\documentclass[]{emulateapj}
\usepackage{graphicx, amsmath, amsthm, amssymb,color}

\def\ba{\begin{eqnarray}}
\def\ea{\end{eqnarray}}
\def\ffrac#1#2{{\textstyle\frac{#1}{#2}}}
\def\in{_{\rm in}}
\def\out{_{\rm out}}

\shorttitle{Warm Jupiters from secular  planet-planet interactions }
\shortauthors{Petrovich \& Tremaine}

\begin{document}

\title{Warm Jupiters from 
secular  planet-planet interactions}
\author{Cristobal Petrovich\altaffilmark{1,2,3} \& Scott Tremaine\altaffilmark{4}}
\altaffiltext{1}{Canadian Institute for Theoretical Astrophysics, University of Toronto, 
60 St George Street, ON M5S 3H8, Canada; cpetrovi@cita.utoronto.ca}
\altaffiltext{2}{Centre for Planetary Sciences, Department of Physical \& 
Environmental Sciences, University of Toronto at Scarborough, Toronto, 
Ontario M1C 1A4, Canada}
\altaffiltext{3}{Department of Astrophysical Sciences, Princeton University, 
Ivy Lane, Princeton, NJ 08544, USA}
\altaffiltext{4}{Institute for Advanced Study, Einstein Drive, Princeton, NJ 08540, 
USA; tremaine@ias.edu}

\begin{abstract}
Most warm Jupiters (gas-giant planets with $0.1~{\rm AU}\lesssim a \lesssim1$ AU) 
have pericenter distances that are too large for
significant orbital migration by tidal friction.  We study the
possibility that the warm Jupiters are undergoing secular eccentricity
oscillations excited by an outer companion (a planet or star) in an
eccentric and/or mutually inclined orbit. In this model the warm
Jupiters migrate periodically, in the high-eccentricity phase of the
oscillation, but are typically observed at lower eccentricities.  
We show that in this model the steady-state eccentricity distribution of the warm
Jupiters is approximately flat, which is consistent with the observed
distribution if we restrict the sample to warm Jupiters with
detected outer planetary companions.  The eccentricity distribution 
of warm Jupiters without companions exhibits a peak at $e\lesssim0.2$ 
that must be explained by a different formation mechanism.  Based on a
population-synthesis study we find that high-eccentricity migration
excited by an outer planetary companion (i) can account for $\sim20\%$
of the warm Jupiters and most of the warm Jupiters with $e\gtrsim
0.4$; (ii) can produce most of the observed population of hot
Jupiters, with a semimajor axis distribution that matches the
observations, but fails to account adequately for $\sim 60\%$ of hot
Jupiters with projected obliquities $\lesssim20^\circ$.  Thus 
$\sim20\%$ of the warm Jupiters and $\sim60\%$ of the hot Jupiters can be
produced by high-eccentricity migration.  We also provide predictions
for the expected mutual inclinations and spin-orbit angles of the
planetary systems with hot and warm Jupiters produced by
high-eccentricity migration.
\end{abstract}                     
                

\section{Introduction}
\label{sec:intro}

We define `warm Jupiters' to be
gas-giant planets with projected mass $M\sin i>0.1$ Jupiter 
masses ($M_J$)
and semimajor axis in the range $0.1~{\rm AU}\leq a \leq1$ AU. As of
September 2015  $\simeq 112$ warm Jupiters had been discovered 
in radial-velocity 
(RV) surveys\footnote{Data from
http://exoplanets.org  \citep{wright11}
and http://exoplanet.hanno-rein.de
    \citep{rein12}.}, compared to 40
planets in the same mass range with $a\leq0.1$ AU  
(commonly called `hot Jupiters')
and $\simeq250$ with $a>1$ AU.
The warm Jupiters have median eccentricity
$\simeq0.24$ and 
median pericenter distance  $a(1-e)\simeq0.33$ AU.
Most ($\sim70\%$)  are in single-planet systems
without any detected companions,
although recent studies suggest that roughly 
half of the warm Jupiters (and also the hot Jupiters) 
have distant ($a=5-20$ AU) planetary-mass companions 
(\citealt{knutson14,bryan16}).

The origin of the warm Jupiters is unexplained. Both hot and warm
Jupiters are thought to have
formed beyond the ice-line at a few AU and 
then migrated inward (e.g., \citealt{bod00,R06}).
The main candidate mechanisms for large-scale orbital migration 
are disk-driven migration and high-eccentricity 
migration.
In the former process, the planet migrates 
by transferring its orbital angular momentum 
to the surrounding gaseous 
protoplanetary disk \citep[e.g.,][]{GT80,W97,bar14}, while
in the latter the planet
attains very high eccentricity by one of a variety 
of mechanisms and then tidal dissipation from the
host star
circularizes the orbit at small semimajor axis  
(e.g., \citealt{RF96,WM03}).

Neither of these mechanisms can easily 
account for the warm Jupiters.
For disk-driven migration it is unclear why 
the migration stopped partway (e.g., \citealt{IL08,mordasini09})
 and why most of the warm Jupiters have relatively high
eccentricities ($e\gtrsim0.2$),
as disk-planet interactions tend to damp rather than excite the
eccentricities \citep{DAA13,Bitsch13}, and eccentricity excitation
through planet-planet gravitational scattering after migration is
ineffective, because scattering at small semimajor axes
generally leads to collisions between the planets \citep{PTR14}.
For high-eccentricity 
migration, the main challenge is to 
explain why most of the warm Jupiters have 
pericenter distances $a(1-e)\gtrsim0.1$ AU at which 
tidal forces are too weak to produce
significant migration.

In this paper, we study a mechanism that overcomes---or at least 
alleviates---this last 
difficulty within the high-eccentricity 
migration scenario. We argue that warm Jupiters periodically
reach smaller pericenter distances than their current ones by exchanging
angular momentum with a distant massive perturber
(e.g., \citealt{WL11,DKS14}).
In this picture, warm Jupiters are observed today
at the low-eccentricity (or large pericenter distance)
phase of  such oscillations, while migration occurs during the
high-eccentricity phase; thus, given enough time,
all warm Jupiters would evolve into hot Jupiters.

One possibility for the distant massive perturber is a stellar or
planetary companion
on a highly inclined orbit. In this case warm Jupiters may exchange
angular momentum with the perturber through Kozai--Lidov 
oscillations (\citealt{kozai}, \citealt{lidov}; see \citealt{naoz16}
for a review).
The stellar-companion scenario is able to produce some but not 
all of the hot Jupiters 
\citep{WM03,FT07,WMR07,naoz12,
petro15a,ASL15,MLL16}, and it can reproduce the orbital 
architecture of the highly eccentric ($e\simeq0.93$) warm
Jupiter HD 80606b (e.g., \citealt{WM03,moutou09,hebrard10}).
However, \citet{petro15a} showed that the stellar-companion scenario 
cannot produce all of the hot Jupiters, and also generally cannot
produce enough warm Jupiters---the migration 
proceeds too fast so the migrating planet 
spends only a small fraction of its lifetime at intermediate
semimajor axes between the cold  ($a\gtrsim1$ AU) and hot Jupiters
($a\lesssim0.1$ AU).
Alternatively, the Kozai--Lidov oscillations can be driven 
by a distant and highly inclined planetary companion
(e.g., \citealt{naoz11,DC14}).

A second possibility is that the warm Jupiters change their
orbital angular momenta through secular perturbations
from a nearly coplanar and eccentric 
perturber such as a distant giant planet 
\citep{li14,petro15b}.
As shown by \citet{petro15b}, this planetary-companion scenario can 
account for most of the hot Jupiters, but it produces
warm Jupiters at a rate that is too low and with eccentricities
that are too large compared to the
observations.

  In this work we examine the effects of secular gravitational
interactions with a distant perturber considering the general case in 
which the perturber
(either a planet or a star) is in an eccentric and/or highly inclined orbit. 
Thus, these perturbations include the Kozai--Lidov mechanism, 
the coplanar and
eccentric case, and possible combinations of these two.
Although our main results are applicable to either 
the planetary- or the binary-companion scenarios, we
focus most our attention on the former scenario because
it seems to be a more promising mechanism for 
explaining the warm Jupiters.
We discuss our results in the
context of the stellar-companion scenario in \S\ref{sec:binary}. 

The frequency and orbital properties of distant ($a\gtrsim 5$ AU) 
companions of the warm Jupiters are largely unknown; the
principal constraints come from linear trends in the radial-velocity
curves of the host star. 
One critical and largely unconstrained orbital element for our
purposes is the mutual inclination
between the planet and the companion orbit $i_{\rm tot}$, which has
been measured only in
a few exceptional examples such as Kepler-419 
b and c ($a\in=0.37$ AU, $a\out=1.7$ AU, 
  $i_{\rm tot}=9^\circ\pm8^\circ$;
\citealt{dawson14})
and Upsilon Andromedae c and d
($a\in=0.86$ AU, $a\out=2.70$ AU, and 
$i_{\rm tot}\simeq30^\circ\pm1^\circ$; 
\citealt{mcarthur10}).
Recently, \citet{DC14}  examined stars hosting warm Jupiters and a
second planet at larger semimajor axis and observed that the sky-plane
apsidal separations $|\omega_{\rm
  out}-\omega\in|$ clustered around $\sim90^\circ$. They argued
that this clustering implies that the mutual inclinations $i_{\rm
  tot}$ oscillate between $\sim35^\circ$ and $65^\circ$.

\section{Prerequisites for the 
formation of warm Jupiters}
\label{sec:conditions}

We first describe the properties of an outer perturber that are needed to
produce low- or moderate-eccentricity warm Jupiters.

The characteristic timescale for 
the eccentricity oscillations due to an outer 
perturber is the Kozai--Lidov timescale
\citep{kiseleva98}: 
\ba
\tau_{\mbox{\tiny{KL}}}=\frac{2P\out^2}
{3\pi P\in}\frac{m_s+m\in+m\out}{m\out}
\left(1-e\out^2\right)^{3/2},
\label{eq:tau_KL}
\ea
where $m_s$, $m\in$, and $m\out$ are the masses of the
host star, the planet, and the outer perturber (the
planetary or stellar companion).
The inner binary has semimajor axis $a\in$ and orbital period
$P\in=2\pi a\in^{3/2}/[G(m_s+m\in)]^{1/2}$, while the
outer binary
has semimajor axis $a\out$, period $P\out=2\pi a_{\rm
  out}^{3/2}/[G(m_s+m\in+m\out)]^{1/2}$ and
eccen\-tricity $e\out$.
In the case of a coplanar and eccentric outer perturber, the period of
the eccentricity
oscillations is longer than the Kozai--Lidov
timescale by a factor of order $e_{\rm
  out}a\in/a\out$ because the oscillations 
arise from octupole rather than quadrupole terms in the gravitational
potential from the outer perturber
\citep{li14,petro15b}. In this case the conditions for eccentricity
oscillation are even more 
stringent than those we give below. 

In order for a migrating warm Jupiter ($m\in \ll m_s$)
to be undergoing 
eccentricity oscillations, we
require at least the following two conditions:

\begin{enumerate}
\item The migration rate must be slow relative
to the oscillation period due to secular perturbations.

More specifically, the secular torque from the companion should be
strong enough that it can change the planet's pericenter distance
before tidal dissipation is able to shrink the semimajor axis.  
In the opposite limit, when the migration is fast, the
planet migrates at roughly constant angular momentum, thereby
producing only high-eccentricity warm Jupiters, with pericenters small
enough that tidal dissipation is always important \citep{petro15a}.

We can quantify the condition for slow migration
by comparing the timescale on which the secular
torque changes the pericenter distance, 
$\tau_{\rm p}=|r_{\rm p}/\dot{r}_{\rm p}|$
with $r_{\rm p}=a\in(1-e\in)$, 
and the migration timescale $\tau_a=|a\in/\dot{a}\in|$.
In the high-eccentricity limit  ($1-e\in\ll1$) which is relevant for
migration, the former timescale can be computed
from $|de\in/dt|\simeq(1-e\in^2)^{1/2}/ \tau_{\mbox{\tiny{KL}}}$
and reduces to $\tau_p\simeq(1-e\in^2)^{1/2} \tau_{\mbox{\tiny{KL}}}$
\footnote{\citet{petro15a} used a different expression to approximate
$\tau_p\simeq(1-e\in^2) \tau_{\mbox{\tiny{KL}}}$, in which case the 
expression for $r_{\rm p,crit}$ in Equation (\ref{eq:r_p}) 
changes only slightly: the coefficient changes from 0.01 AU to
$0.009$ AU, the exponent of $a\in$ changes from 3 to $5/2$,
and the exponent of the square brackets changes from $1/7$ to $2/13$.
The results in \citet{petro15a} are unchanged if he used the expression
$r_{\rm p,crit}$ from Equation (\ref{eq:r_p}). 
},
while the latter is \citep{petro15a}: 
\begin{equation}
\qquad\tau_a\simeq\frac{2^{25/2}}{34749}(1-e\in)^{15/2} \left(\frac{a\in}{R_{\rm p}}\right)^8
\frac{t_{\rm V,p}}{(1+k_{\rm p})^2}\left(\frac{m_{\rm
      in}}{m_s}\right)^2
\end{equation}
where  $k_{\rm p}\simeq 0.5$ is the Love number of the planet and 
$t_{\rm V,p}$ is the viscous time of the planet.
The critical pericenter distance $r_{\rm p,crit}$ at which $\tau_{\rm
  p}=\tau_a$ is:
 \begin{align}
\qquad & r_{\rm p,crit}\simeq 0.01 \mbox{ AU} \left[ \left(\frac{m_s}{M_\odot}\right)^\ffrac{5}{2}
\left(\frac{M_J}{m\in}\right)^2\left(\frac{0.1\mbox{ yr}}{t_{\rm V,p}}\right) 
 \right.\nonumber\\
&\times\left.  \left(\frac{R\in}{R_J}\right)^{8}\!
\left(\frac{1\mbox{ AU}}{a\in}\right)^\ffrac{5}{2}\!	
  \left(\frac{a\out\sqrt{1-e\out^2}}{10\mbox{ AU}}\right)^3\!	
  \left(\frac{M_J}{m\out}\right)
\right]^\ffrac{1}{7}\!\! \!\!\nonumber\\
\label{eq:r_p}
\end{align}
Here $R\in$ is the radius of the planet and $R_J$ is Jupiter's radius.
For $r_{\rm p}<r_{\rm p,crit}$ we have $\tau_a<\tau_{\rm p}$
and the planet migrates fast, at roughly constant angular
  momentum. On the contrary, if $r_{\rm p}$ is even slightly larger
than $r_{\rm p,crit}$ the strong dependence of tidal dissipation
on $r_{\rm p}$ ensures that $\tau_a\gg\tau_{\rm p}$
and the planet undergoes slow migration.

\item The pericenter precession rate due to extra forces must be
slow relative to the precession due to secular perturbations.

If the inner planet is undergoing slow migration
and its migration timescale is less than the age of the system, 
then the eccentricity oscillations will continue down to a critical
semimajor axis $a_{\rm in, crit}$ at which extra forces make
the inner orbit precess fast enough that the eccentricity oscillations
are quenched.
As shown by \citet{SKDT12} and \citet{DKS14}, if the dominant 
precession force
is general relativity, as is often the case, this critical
semimajor axis is 
\ba
a_{\rm in,  crit}&\simeq& 0.5 \mbox{ AU} \left(\frac{m_s}{M_\odot}\right)^{4/7}
  \left(\frac{M_J}{m\out}\right)^{2/7}\nonumber\\
&\times&  \left(\frac{a\out\sqrt{1-e\out^2}}{10\mbox{ AU}}\right)^{6/7}.
  \label{eq:acrit}
\ea
Thus, the eccentricity  
oscillations are constrained to $a\in\gtrsim a_{\rm in,  crit}$.

\end{enumerate}

Based on the conditions above, the production of warm Jupiters
is facilitated for planets with less efficient dissipative properties
(larger $t_{\rm V,p}$ and/or smaller $R\in$) and outer 
perturbers with higher masses
$m\out$ and smaller semimajor axes $a\out$.


\begin{figure*}[t!]
   \centering
      \includegraphics[width=15cm]{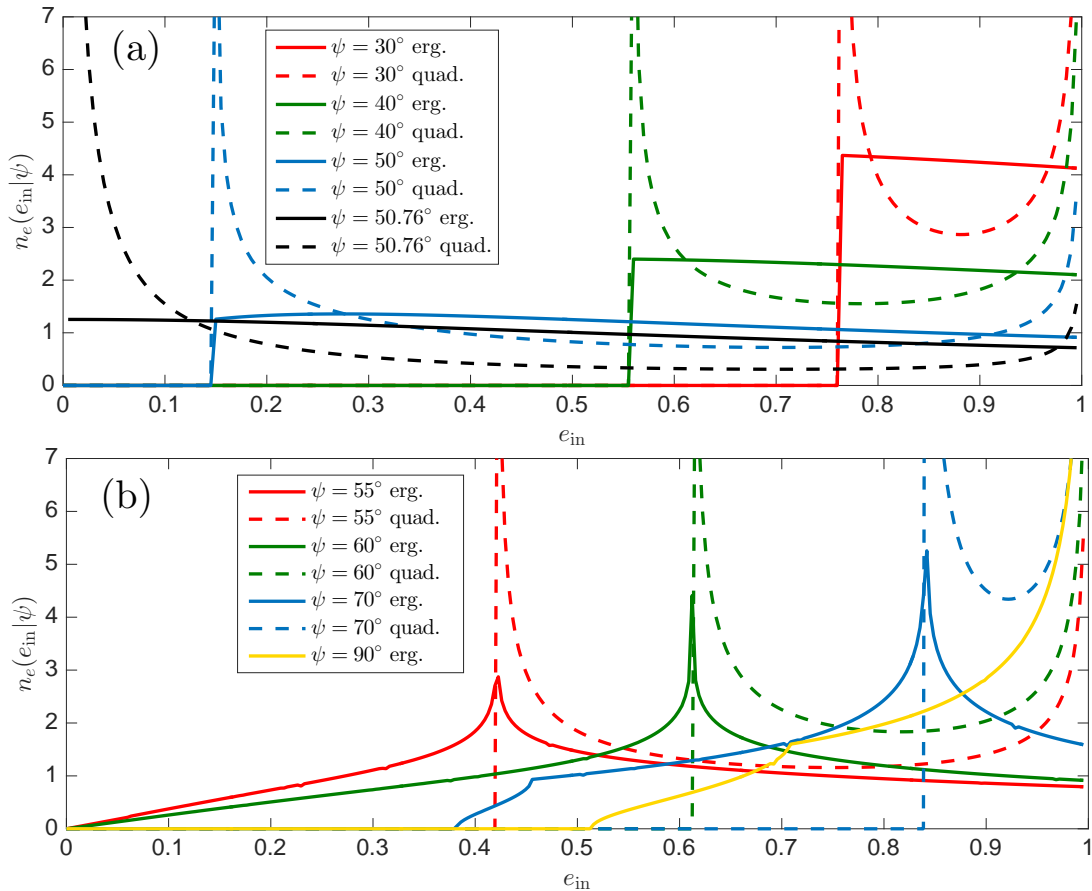}
  \caption[Time-averaged eccentricity distributions 
  for a migrating planet in the quadrupole and ergodic
  approximations] 
  {Time-averaged eccentricity distributions for a migrating planet,
    $n_e(e\in|\psi)$, for different values of the polar angle of the
    eccentricity vector $\psi$ at $e\in=1$.  The dashed lines show the
    result from the quadrupole approximation in Equation
    (\ref{eq:n_e_psi}), and the solid lines show the result from the
    ergodic approximation in Equation (\ref{eq:n_e_psi_ellip}).  {\it
      Panel (a):} $\psi\leq50.76^\circ$.  {\it
      Panel (b):} $\psi>50.76^\circ$.  }
\label{fig:n_e_erg_quad}
\end{figure*}

\section{Eccentricity distribution for a migrating
planet}
\label{sec:e_dist}

We now study the time-averaged eccentricity distribution of a planet
that undergoes migration
due to secular gravitational
interactions from an outer perturber. 
We assume that migration is due to tides from the host star 
and thus can only happen if the planet reaches 
$e\in\simeq1$ at some point of its orbital evolution.

\subsection{Preliminary definitions}
\label{ref:definitions}

We use the notation from \citet{petro15a}, in 
which the variables are the eccentricity vectors ${\bf e}\in$
and ${\bf e}\out$, and the orbital angular momentum vectors
${\bf h}\in$ and ${\bf h}\out$, all defined in the Jacobi reference
frame (see also \citealt{liu15}).

The doubly time-averaged interaction potential up to octupole
order $\phi_{\rm oct}$ can be written in dimensionless 
form\footnote{Note that this interaction 
potential $\phi_{\rm oct}$ in \citet{petro15a} has positive energy, 
contrary to our convention
here.}
as:
\ba
\tilde\phi_{\rm{oct}}&=&\frac{\phi_{\rm oct}}{\phi_0}=
-\frac{1}{(1-e\out^2)^{3/2}} 
\Big[ \ffrac{1}{2} \left(1-e\in^2\right)
\big({\bf \hat{h}}\in\cdot {\bf \hat{h}}\out\big)^2  
\nonumber\\
&+& \left(e\in^2 -\ffrac{1}{6}\right)-
\ffrac{5}{2}  \big({\bf e}\in\cdot {\bf \hat{h}}\out\big)^2   \Big]\nonumber\\
&-&\frac{\tilde\epsilon_{\rm{oct}}}{\left(1-e\out^2\right)^{3/2}} 
\Big\{ 
 \big({\bf e}\in\cdot {\bf \hat{e}}\out\big)
 \Big[ \big(\ffrac{1}{5}-\ffrac{8}{5}e\in^2\big)
\nonumber\\
 &-& \left(1-e\in^2\right)
 \big({\bf \hat{h}}\in\cdot {\bf \hat{h}}\out\big)^2+
 7  \big({\bf e}\in\cdot {\bf \hat{h}}\out\big)^2
\Big]
	 \nonumber\\
	 \label{eq:phi_oct}
&-&	
2\left(1-e\in^2\right)\big({\bf \hat{h}}\in\cdot {\bf \hat{h}}\out\big)
\big({\bf e}\in\cdot {\bf \hat{h}}\out\big)
 \big({\bf \hat{h}}\in\cdot {\bf \hat{e}}\out\big)
\Big\},\nonumber\\
\ea
where
   \ba
   \phi_0&=&\frac{3Gm\in m\out a\in^2}{4a\out^3}, 
\label{eq:phi_0}\\
   \tilde\epsilon_{\rm{oct}}&=&
   \ffrac{25}{16}\frac{a\in}{a\out}\frac{e\out}{(1-e\out^2)}.
   \label{eq:eoct}
\ea

For a radial orbit ($e\in=1$) the potential in Equation 
(\ref{eq:phi_oct}) reduces to
\ba
\tilde\phi_{\rm oct}&=&-
\frac{1}{\left(1-e\out^2\right)^{3/2}}\Big\{
\ffrac{5}{6} -\ffrac{5}{2}  ({\bf \hat{e}}\in\cdot {\bf \hat{h}}\out)^2 
\nonumber\\
&+&7\tilde\epsilon_{\rm{oct}}
 ({\bf \hat{e}}\in\cdot {\bf \hat{e}}\out)\left[ 
   ({\bf \hat{e}}\in\cdot {\bf \hat{h}}\out)^2-\ffrac{1}{5}\right]
   \Big\},
   \label{eq:phi_oct_e1}
\ea
and we can write this potential in terms of the
spherical coordinates\footnote{\label{footpsi}
Unlike the orbital inclination $i\in$, the argument of pericenter
$\omega\in$, and 
the longitude of the ascending node $\Omega\in$,  
the polar and azimuthal angles
of the eccentricity vector are well-defined for
radial orbits, $e\in=1$ (e.g., \citealt{tremaine2001}). One can relate 
the spherical angles of the eccentricity vector to the more 
commonly used Delaunay orbital angles as:
$\cos\psi=\sin\omega\in\sin i\in$ and 
$\sin\psi\cos(\phi-\Omega\in)=\cos\omega\in$.} 
of the eccentricity vector
at $e\in=1$ as:
\ba
\tilde\phi_{\rm oct}&=&-
\frac{1}{\left(1-e\out^2\right)^{3/2}} \Big\{\ffrac{5}{6} -\ffrac{5}{2}  \cos^2\psi 
\nonumber\\
&+&7\tilde\epsilon_{\rm{oct}}
\cos\phi\sin\psi\left[ 
   \cos^2\psi-\ffrac{1}{5}\right]
   \Big\},
   \label{eq:phi_oct_angle}
\ea
where $\psi$  is the polar angle and $\phi$
is the longitude of the eccentricity vector in the
orthogonal basis 
$({\bf \hat{e}}\out,{\bf \hat{h}}\out\times{\bf \hat{e}}\out
,{\bf \hat{h}}\out)$.

\subsection{Quadrupole approximation}
\label{sec:e_KL}

In the Appendix, we derive the time-averaged 
eccentricity distribution $n_e(e\in)$ of a test particle
($m\in\ll m_s,m\out$) undergoing secular eccentricity oscillations 
due to an external quadrupole potential ($\tilde\epsilon_{\rm oct}=0$ in 
Equation \ref{eq:phi_oct}). The derivation assumes that $e\in=1$ 
at some phase of the
oscillation.

In particular, in Equation (\ref{eq:n_e_psi}) 
we give  $n_e(e\in|\psi)$, where $\psi$ is the polar angle of the eccentricity 
vector  when $e\in=1$.
From the denominator in that equation it is straightforward 
to show that this distribution is defined if and only if  
$e_{\rm in, min}\leq e\in\leq1$,
where 
\begin{equation}
  e_{\rm in, min}=\begin{cases}
    \left(1-\tfrac{5}{2}\cos^2\psi\right)^{1/2}, & \text{if $|\cos\psi|\leq\sqrt{2/5}$}\\
    \left(\tfrac{5}{3}\cos^2\psi-\tfrac{2}{3}\right)^{1/2}, & \text{otherwise}.
  \end{cases}
  \label{eq:e_min}
\end{equation}

In Figure \ref{fig:n_e_erg_quad} we show
$n_e(e\in|\psi)$ from Equation (\ref{eq:n_e_psi}) for 
different values of $\psi$ (dashed lines, labeled `quad').
We observe that $n_e(e\in|\psi)$ always diverges as
$e\in\to 1$ and at the
minimum eccentricity given by Equation  (\ref{eq:e_min}).
This behavior is expected because the eccentricity oscillations 
due to the  Kozai--Lidov mechanism  have turning points
at these eccentricities.
Also, consistent with our Equation  (\ref{eq:e_min}), we observe 
that low eccentricities can only be achieved
when $|\cos\psi|\simeq\sqrt{2/5}$ ($\psi\simeq50.76^\circ$ or
$129.23^\circ$)\footnote{
In the quadrupole
approximation, a test particle reaches a maximum eccentricity 
$e\in\simeq1$ at $\sin^2 \omega\in=1$, implying that the 
minimum inclination is $\sin^2 i_{\rm min}=\frac{2}{5}$ 
($i_{\rm min}=39.2^\circ$ and $140.7^\circ$) 
if $e\in$ passes through 0
(see the derivation in \S\ref{sec:itot}). 
From the relation 
$\cos\psi=\sin\omega\in\sin i\in$ (footnote \ref{footpsi})
we observe that trajectories connecting $e\in=0$
with $e\in=1$  satisfy 
$\cos( \psi+i_{\rm min})=0$ so
$\psi+i_{\rm min}=90^\circ$ or $270^\circ$.}.

In summary, in the approximations used here (quadrupole potential and
planet of negligible mass) a
migrating planet  spends most of its time at eccentricities near unity
or near a minimum value that depends only on 
the polar angle of the eccentricity vector at $e\in=1$
(Eq.\  \ref{eq:e_min}).
The planet reaches low eccentricities during the Kozai--Lidov cycle if and
only if this angle is $\simeq50.76^\circ$ or $129.23^\circ$.

\subsection{Ergodic approximation}
\label{sec:e_ergodic}

As described in the previous subsection, in the quadrupole
approximation the secular evolution of the 
inner planet
 has one degree of freedom and is integrable. 
This is not generally the case 
when this approximation is dropped
and finding an approximate analytic description of the steady-state 
eccentricity distribution 
then becomes more challenging.

In this section, we approach this problem using the ergodic
approximation: we assume that the planetary orbits randomly 
populate all the available phase space allowed by conservation of
energy. 
We shall test this hypothesis using numerical integrations
in \S\ref{sec:compa_erg}.
We recall that the derivations in the Appendix used in this section
are valid in the test particle regime ($m\in\ll m_s,m\out$).

From Equation (\ref{eq:phi_oct_angle}) we observe that
in the quadrupole approximation ($\tilde\epsilon_{\rm oct}=0$)
the energy of a migrating planet is uniquely determined
by  $\psi$---the polar angle of the eccentricity vector 
at $e\in=1$. Therefore, given a value of
$\psi$ we can determine a unique
time-averaged eccentricity distribution. In the ergodic approximation, 
the component of angular momentum normal to the perturber orbit,
$H\in$, is uniformly distributed between $\pm L\in$ so we may 
integrate Equation (\ref{eq:n_e}) over $H_0$ and use Equation
(\ref{eq:theta_psi}) to obtain
\ba
n_e(e\in|\psi)&\propto&\frac{e\in}{
(1+4e\in^2)^{1/2}}\times\nonumber\\
&&\Re\left\{\int_{0}^1\frac{du}{\left[(A+u^2)(B-u^2)\right]^{1/2}}
\right\},
  \label{eq:n_e_psi_ellip}
  \ea
  where
  \ba
  A&=&2e\in^2-2+5\cos^2\psi,\label{eq:A}\\
  B&=&\frac{\left(1-e\in^2\right)\left( 2+ 3e\in^2-5\cos^2\psi\right)
  }{1+4e\in^2},\label{eq:B}
\ea
and $\Re$ is the real part.
We note that this integral can be expressed in terms of 
elliptic integrals, but the expression for arbitrary $\psi$
is somewhat cumbersome and uninformative.

In Figure \ref{fig:n_e_erg_quad} we show
$n_e(e\in|\psi)$ from Equation (\ref{eq:n_e_psi_ellip}) for 
different values of $\psi$ (solid lines).
We observe that for $\psi\leq50.76^\circ$ (top panel) the distribution is 
relatively flat and restricted to high eccentricity, with a 
minimum $e_{\rm in,min}$ given by 
Equation (\ref{eq:e_min}).
This flat profile contrasts with the profile given by the 
quadrupole approximation, 
which exhibits strong peaks at $e=1$ and the minimum eccentricity
$e_{\rm in,min}$. 

\begin{figure*}[t!]
   \centering
  \includegraphics[width=18cm]{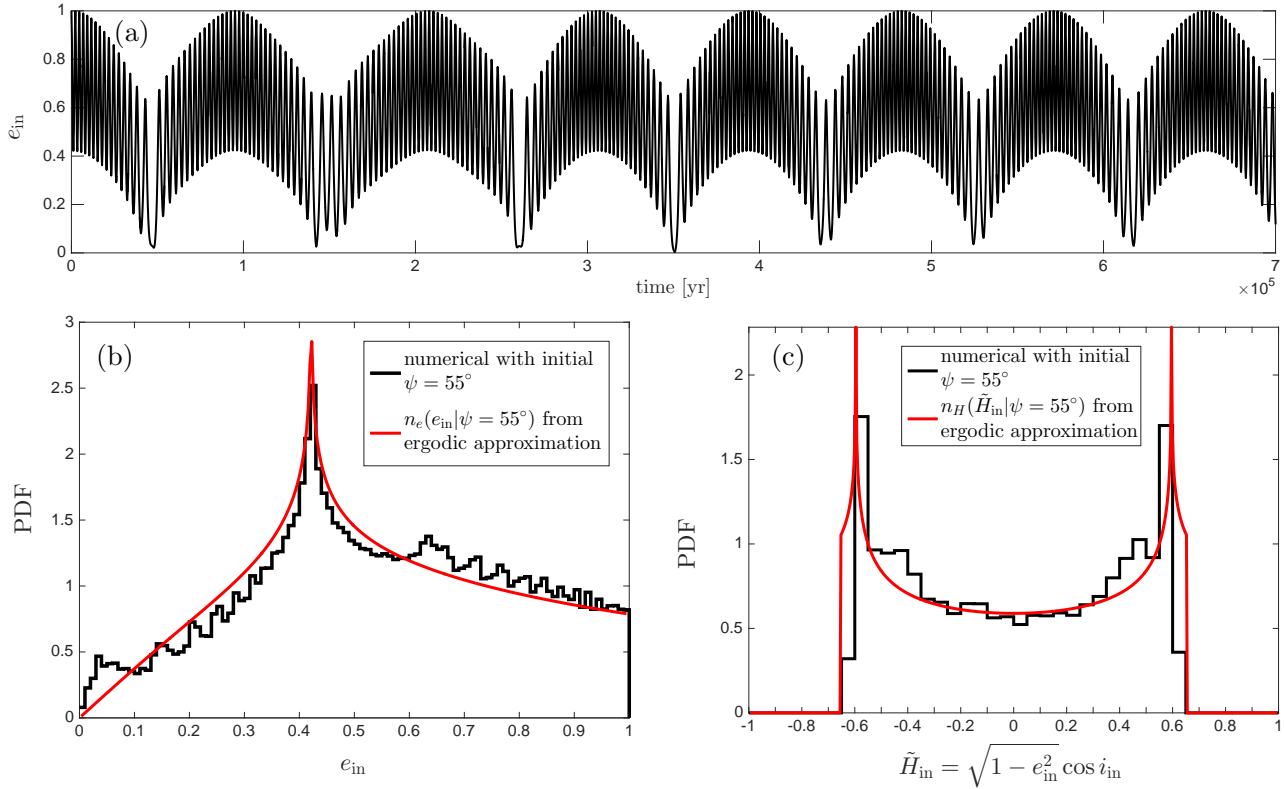}
  \caption[Example of the secular evolution of a triple system
  and comparison of the eccentricity distribution
  with the ergodic approximation]{Secular evolution of a triple system
  with parameters  $a\in=1$ AU,  $a\out=10$ AU, 
  $e\out=0.5$,
  $m_s=1M_\odot$, $m\in=1M_J$,
   and $m\out=0.1M_\odot$.
  The inner planet is initially placed in a nearly radial orbit with
  $e\in=0.9999$ and an eccentricity vector 
having polar and azimuthal angles $\psi=55^\circ$ and
$\phi=90^\circ$. The equator of the reference frame coincides with the orbital
plane of the outer body and the $x$-axis of the frame coincides with
its eccentricity vector. 
{\it Panel (a)}: evolution of the eccentricity $e\in$ of the inner orbit.
{\it Panel (b):} eccentricity distribution (solid black line) averaged
over 1000 Kozai--Lidov timescales as defined by Equation (\ref{eq:tau_KL}) ($\simeq1.4$ Myr). 
We also show the
distribution $n_e(e\in|\psi=55^\circ)$ from
the ergodic approximation of Equation (\ref{eq:n_e_psi_ellip}) 
(solid red line).
{\it Panel (c):} distribution of $\tilde{H}\in$ over 1000 Kozai--Lidov timescales.
The solid red line shows the distribution  $n_H(\tilde{H}\in|\psi=55^\circ)$
from the ergodic approximation, (Eq.\ \ref{eq:nH}).
}
\label{fig:compa55}
\end{figure*}  

For $\psi>50.76^\circ$ the behavior of $n_e(e\in|\psi)$
changes in nature: the distribution is peaked
at an intermediate eccentricity. 
The position of this peak coincides with the
minimum eccentricity in the quadrupole approximation
(dashed lines) from Equation (\ref{eq:e_min}). The range of eccentricities 
is significantly wider  than in the quadrupole approximation.
In particular,  small eccentricities can occur for a wide range in $\psi$.
We can estimate this range from Equation (\ref{eq:n_e}) by observing
that at low eccentricities the two factors in the square root are only
positive for some value of $H_0/L$ if $\theta_0$ lies between $-1$
and 0; and from equation (\ref{eq:theta_psi}) this requires that 
$|\cos\psi|$ is in the range $[\sqrt{1/5},\sqrt{2/5}]$ ($\psi$ in the
range $[50^\circ.76,63^\circ.43]$),
consistent with Figure \ref{fig:n_e_erg_quad} (solid lines
with $\psi=\{50.76^\circ,55^\circ,60^\circ\}$).

From Figure \ref{fig:n_e_erg_quad} we observe that for all 
values of $\psi$, except for the limiting case $\psi=90^\circ$,
the eccentricity distribution predicted by the 
ergodic approximation is a decreasing function of $e\in$ 
as it approaches unity.
This behavior is different from the quadrupole approximation, in which
the eccentricity diverges at $e\in=1$.
All else being equal, the migration speed 
is determined by the fraction of time that the planet spends
at $e\in\simeq1$. Since this fraction is significantly lower in the
ergodic approximation (compare solid and 
dashed lines at $e\in=1$), we expect that the migration speed
predicted by the ergodic approximation is much slower than 
that from the quadrupole approximation.

In summary, the ergodic approximation predicts two families
of eccentricity distributions, depending on the polar angle $\psi$ of
the eccentricity vector when the planet first starts to migrate
($e\in=1$): (i) for $\psi\leq50.76^\circ$ the distribution
is flat and allows for the same range of eccentricities as that in the 
quadrupole approximation; 
(ii) for $\psi>50.76^\circ$ the distribution is peaked at an intermediate 
eccentricity and allows for a much wider range of eccentricities 
than predicted by the quadrupole 
approximation.
Unlike the distribution in the quadrupole approximation, 
both families of eccentricity distributions are decreasing functions of
$e\in$ as it approaches unity, implying slower migration
rates.

\begin{figure}[h!]
   \centering
  \includegraphics[width=8.3cm]{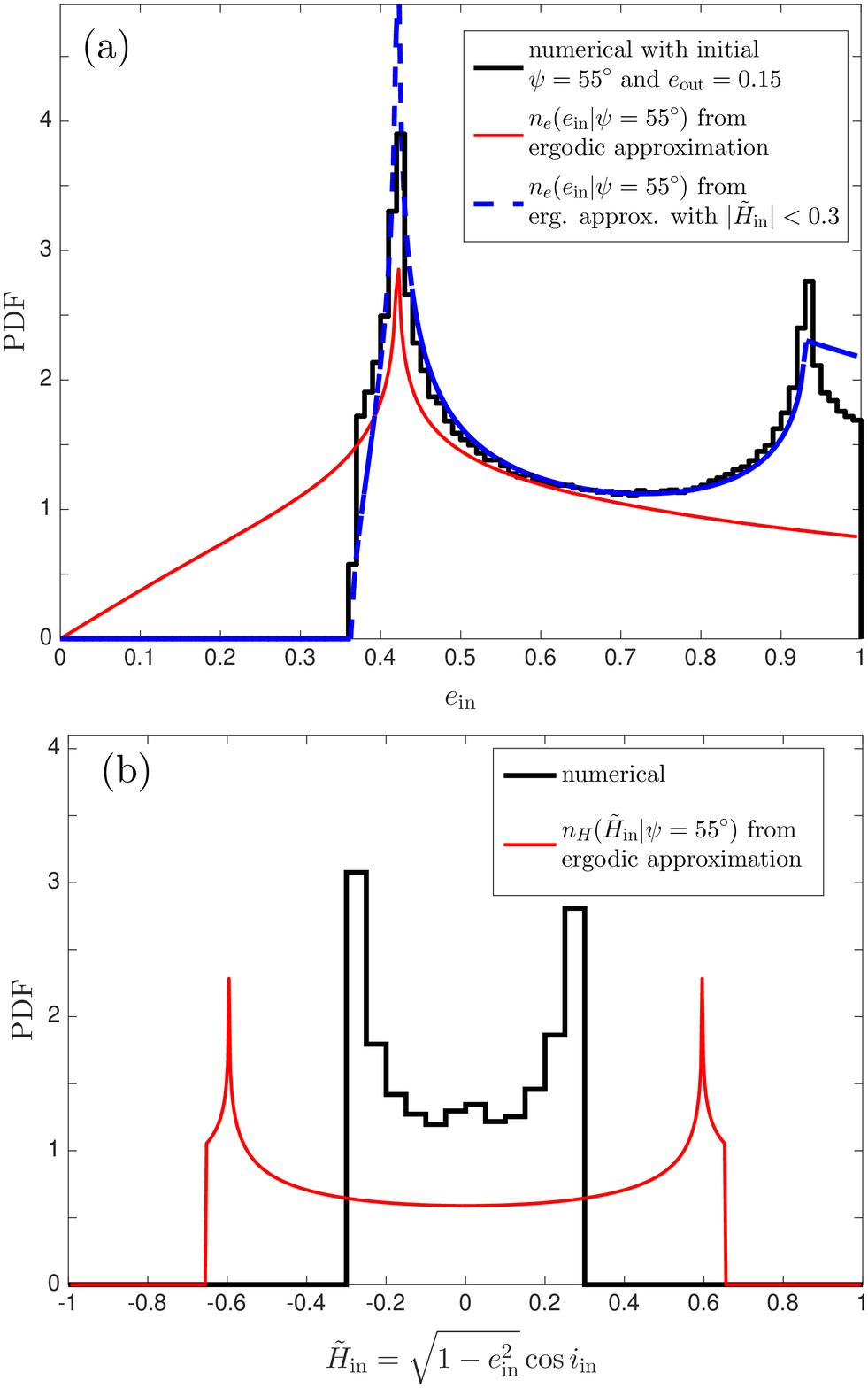}
    \caption[Example of the secular evolution of a triple system
  and comparison of the eccentricity distribution
  with the quadrupole and ergodic approximations]
{The same as panels b and c of Figure \ref{fig:compa55},
  but for a lower eccentricity of the perturber, $e\out=0.15$ 
  compared to 0.5. In both panels the solid black line shows the
  results of a numerical integration of the secular equations of motion.
{\it Panel (a):} the solid red line shows the distribution
$n_e(e\in|\psi=55^\circ)$ expected from
the ergodic approximation (Eq.\ \ref{eq:n_e_psi_ellip}), which does
not agree with the numerical integration. The dashed blue line shows
the distribution obtained if the domain 
of the integral in Equation (\ref{eq:n_e_psi_ellip}) is restricted to
$u=|\tilde H\in|=|(1-e\in^2)^{1/2}\cos i\in|\leq0.3$.
{\it Panel (b):} the solid red line shows the distribution of
$n_H(\tilde H\in|\psi=55^\circ)$ 
expected from the ergodic approximation (Eq.\ \ref{eq:nH}).
}
\label{fig:compa_e015}
\end{figure}

\begin{figure}[h!]
   \centering
  \includegraphics[width=8.3cm]{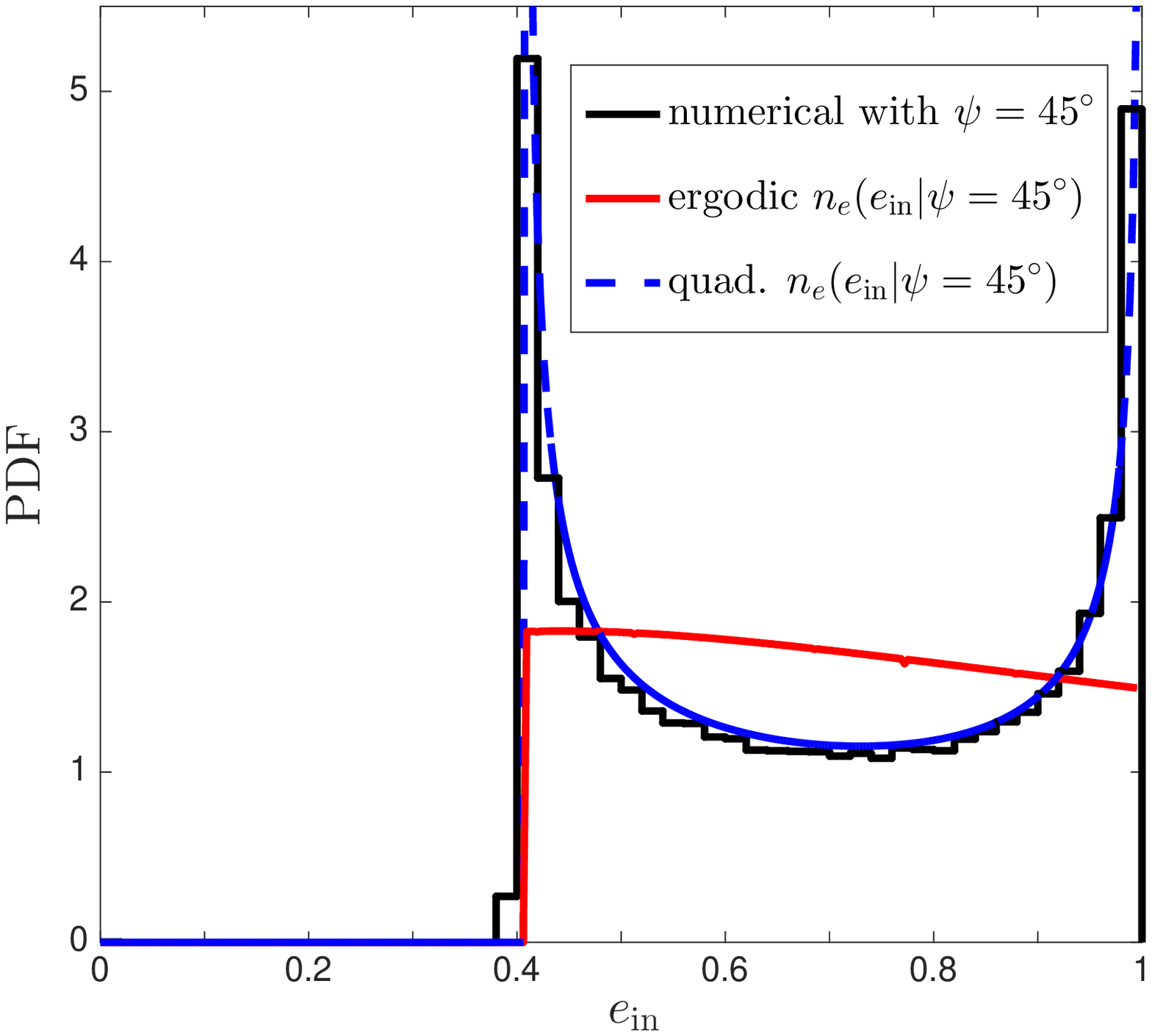}
  \caption[Example of the secular evolution of a triple system
  and comparison of the eccentricity distribution
  with the quadrupole and ergodic approximations]
  {The same as panel b of Figure \ref{fig:compa55}, but with the polar
    angle of the initial eccentricity vector
    $\psi=45^\circ$. 
    We show the distribution $n(e\in|\psi=45^\circ)$ from the ergodic
    approximation (Eq.\ \ref{eq:n_e_psi_ellip}) as the solid red line
    and the distribution using the quadrupole approximation (Eq.\
    \ref{eq:n_e_psi}) as the blue dashed line.  In this example, the
    momentum variable in the numerical integration is found to be
    restricted to the range $|\tilde H\in|\lesssim 0.1$, while the
    ergodic approximation predicts that $|\tilde H\in|$ is flat in
    $[0,0.41]$.  Therefore, in this example
    the quadrupole approximation is expected
    to be more accurate than the ergodic approximation.  }
\label{fig:compa45}
\end{figure}  

\subsection{Comparison with numerical integrations}
\label{sec:compa_erg}

We now compare the analytic results for the eccentricity distribution
obtained using the ergodic approximation in \S\ref{sec:e_ergodic}
with numerical integrations. These involve solving the secular equations 
of motion including the gravitational potential of the external 
perturber up to the 
octupolar moment (up to $a\in^3/a\out^4$). The
equations of motion are given in \citet{petro15a},
after ignoring the non-Keplerian interactions
and assuming that all bodies are point masses
(no tidal disruptions or collisions).

In Figure \ref{fig:compa55} we show the 
evolution of $e\in$ (panel a) for one example that starts 
with an eccentricity vector having magnitude $e\in=1$, polar angle
$\psi=55^\circ$, and azimuth $\phi=90^\circ$\,\footnote{ For our choice 
of $\phi=90^\circ$, 
the energy  $\tilde\phi_{\rm oct}$ in Equation (\ref{eq:phi_oct_e1})
coincides with that in the quadrupole approximation, meaning
that the energy in the ergodic approximation (quadrupole-level) and that
in the numerical integration (octupole-level) are the same.}.
We see that the evolution of $e\in$ 
has a complicated quasi-periodic pattern.

In panel b we show the time-averaged distribution of $e\in$ from the
numerical integration (solid black line) and compare this with our analytic
distribution $n_e(e\in|\psi)$ with $\psi=55^\circ$, given by 
Equation (\ref{eq:n_e_psi_ellip})
based on the ergodic approximation (solid red line).
We observe that the analytic expression describes 
the position of the peak and the overall shape 
of the distribution derived from the simulation
reasonably well.

In panel c of Figure \ref{fig:compa55}, we show the distribution
of $\tilde H\in\equiv H\in/L\in=(1-e\in^2)^{1/2}\cos i\in$
to assess whether the ergodic hypothesis is satisfied and
the planetary orbit randomly populates all the available phase space. 
We compare this distribution with that from 
the ergodic approximation (dashed red line) 
by integrating Equation (\ref{eq:n_e}) over $e\in$ and
using Equation (\ref{eq:theta_psi}) to obtain
\ba
n_H(\tilde{H}\in|\psi)\propto \Re
\int_{0}^1{\frac{e\in de\in}{
\left[(1+4e\in^2)(A+\tilde{H}_{\rm in}^2)(B-\tilde{H}_{\rm in}^2)\right]^{1/2}}},
\label{eq:nH}\nonumber\\
\ea
where $A(e\in,\psi)$ and $B(e\in,\psi)$ are given by
Equations (\ref{eq:A}) and (\ref{eq:B}).
We observe that $n_H(\tilde H\in|\psi=55^\circ)$ reproduces
the distribution of $\tilde H\in$ from the numerical integration
reasonably well.

We now repeat the calculations shown in Figure \ref{fig:compa55}, but
for a simulation with a lower eccentricity of the perturber,
$e\out=0.15$ instead of 0.5, which reduces the strength of the
octupole relative to the quadrupole (lower $\tilde\epsilon_{\rm oct}$
in Eq.\ \ref{eq:eoct}). We show the resulting eccentricity
distribution as the black curve in panel a of Figure
\ref{fig:compa_e015} and observe that this distribution is not
reproduced by the ergodic model (red curve, computed using Eq.\
\ref{eq:n_e_psi_ellip}).  We also show a blue dashed curve
representing the result if the integral in Equation
(\ref{eq:n_e_psi_ellip}) is restricted to the domain $u=|\tilde
H\in|\leq0.3$; this {\em ad hoc} fix dramatically improves the
agreement between the analytic theory and the results from the
numerical integrations. We speculate that the lower value of $e\out$ in
this example has reduced the strength of the octupole-level
perturbations to the point that they can drive the momentum
variable $|\tilde H\in|$ to sample only a restricted part of the 
available region of phase space.

We now repeat the calculation shown
in Figure \ref{fig:compa55} once again, but for a simulation with initial 
condition $\psi=45^\circ$ instead of $\psi=55^\circ$.
We show the resulting eccentricity distribution
in Figure \ref{fig:compa45}  and observe that it is not reproduced 
by the ergodic model.
This disagreement is not surprising because the inner planet in 
this example evolves only through the range 
$|\tilde{H}\in|\lesssim0.1$, while the ergodic hypothesis predicts
that the orbit should fill 
the available phase-space ($|\tilde{H}\in|$ in $[0,0.41]$ from
Eq.\ \ref{eq:n_e}).
In fact, we observe that the quadrupole approximation (solid
red line), which is equivalent to setting  $\tilde{H}\in=0$, fits
the eccentricity distribution in the simulation much better.

We have tested our analytical results with various other examples
and found that when the momentum coordinate $|\tilde{H}\in|$
reaches values close to the maximum allowed by the
conservation of energy (maximum $|\tilde{H}\in|$ such $n_e>0$
in Eq.\ \ref{eq:n_e}), then 
the ergodic approximation gives a
fair description of the eccentricity distribution. If $|\tilde{H}\in|$
remains small during the integration (say $\lesssim0.1$), then the quadrupole 
approximation works much better. 
There are intermediate cases that can be modeled better just 
by limiting the domain of the integral in  Equation (\ref{eq:n_e_psi_ellip})
to a maximum value of $|\tilde{H}\in|$, as we have done in
panel a of Figure \ref{fig:compa_e015} (dashed blue line).

In summary, the ergodic approximation reproduces the results
from numerical simulations reasonably well provided that the specific
angular momentum coordinate $\tilde{H}\in$ fills
a significant part of its available phase-space. 
When this is not the case, then either using the quadrupole approximation
or restricting the domain of $\tilde{H}\in$ in the ergodic
approximation can reproduce the eccentricity distribution from
simulations much better. 

\begin{figure*}
   \centering
  \includegraphics[width=15cm]{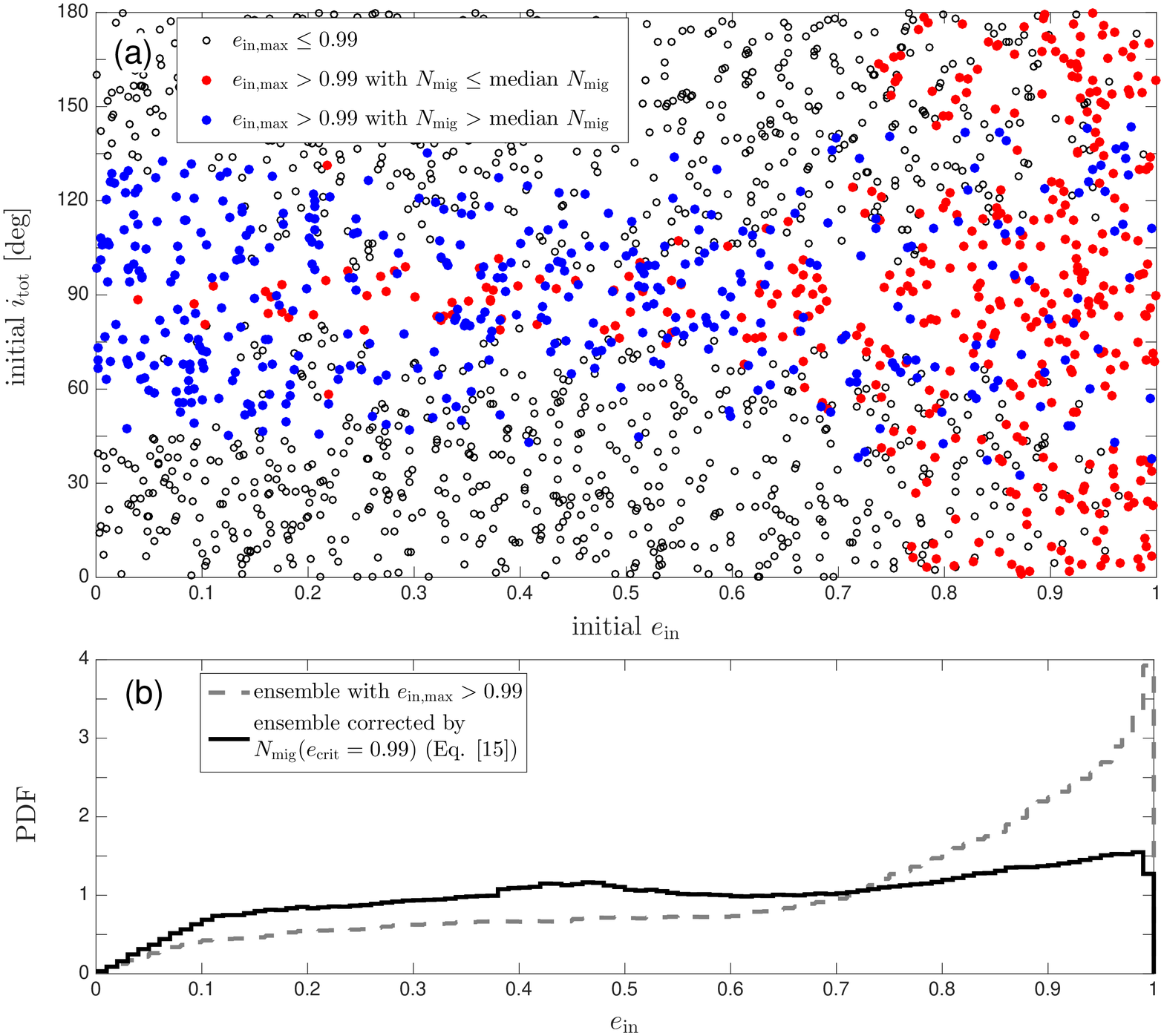}
  \caption{Numerical integration of 2000 triple systems with
    parameters $a\out/a\in=10$, $e\out=0.5$, $m_s=1M_\odot$,
    $m\in=1M_J$, and $m\out=0.1M_\odot$.  The initial conditions are
    chosen from uniform distributions in $\omega\in-\omega\out$,
    $\Omega\in-\Omega\out$ , $e\in$, and $i_{\rm tot}$.  {\it Panel
      (a):} initial mutual inclinations $i_{\rm tot}$ and planet
    eccentricities $e\in$.  The systems that migrate (maximum of
    $e\in>0.99$) are shown as  blue or red circles; those with 
    $N_{\rm mig}(e_{\rm crit}=0.99)$ from Equation
    (\ref{eq:Nmig}) smaller than the median (i.e., systems with faster migration than
    the median) are shown as filled red circles, while
    the systems with $N_{\rm mig}(e_{\rm crit}=0.99)$ larger than the
    median (slower migration) are indicated with filled blue circles.
    Other systems are represented by open black circles.  {\it Panel
      (b):} time-averaged (over 1000 Kozai--Lidov timescales)
    eccentricity distribution for the systems that reach $e\in>0.99$
    (gray dashed line).  The solid black line shows the same
    distribution after correcting for the rate of migration as in
    Equation (\ref{eq:ne_ss}).  }
\label{fig:ss_e}
\end{figure*}

\section{Steady-state eccentricity distribution of 
warm Jupiters}
\label{sec:ss_e}

In this section we calculate the steady-state eccentricity
distribution of warm Jupiters, taking into account the time that these
planets spend undergoing migration.

The migration speed of a planet is mostly determined by the time
it spends at small pericenter distances (high eccentricities). 
Since tidal dissipation
is a very steep function of this pericenter distance, we can 
approximate the migration speed by a step function
$\Theta$ as
\ba
\dot{a}\in&=&v_{\rm mig}\Theta[r_{\rm p,crit}-a\in(1-e\in)]\nonumber\\
&=&v_{\rm mig}\Theta\left[e\in-e_{\rm crit}(a\in)\right],
\ea
where $r_{\rm p,crit}$ and $e_{\rm crit}(a\in)\equiv
       1-r_{\rm p,crit}/a\in$ are the critical pericenter 
distance  and eccentricity at which tidal dissipation is efficient enough 
to cause migration and $v_{\rm mig}=Ca\in^{1/2}$ where $C$ is 
a function of the physical 
properties of the star and the planet (viscous times, 
Love number, radii, etc)\footnote{The dependence of $v_{\rm mig}$ on
  $a\in$ arises as follows. At each pericenter passage a planet on
  a high-eccentricity orbit with a given pericenter distance loses  a
  fixed energy $\Delta E$. Thus $\dot E\propto \Delta E/P$ where
  $P\propto a^{3/2}$ is the orbital period. Since $E\propto a^{-1}$ we
  find $\dot a\propto a^{1/2}$.}.
Then, all else being equal and assuming that our choice of
$e_{\rm crit}$ allows for significant migration,
 the steady-state number of
migrating planets ($e\in>e_{\rm crit}$ at some point of the 
eccentricity evolution) as a function of the critical eccentricity is 
\ba
N_{\rm mig}(e_{\rm crit})\propto \frac{\Delta t}{\int_0^{\Delta t} 
dt\Theta\left[e\in(t)-e_{\rm crit}\right]},
\label{eq:Nmig}
\ea
where the time interval $\Delta t$ is much longer than the 
Kozai--Lidov timescale $\tau_{\mbox{\tiny{KL}}}$ in Equation
(\ref{eq:tau_KL}).

Thus, for an ensemble of $i=1,...,N$ planetary systems that 
have time-averaged eccentricity distributions $n_e^i(e\in)$, the 
expected steady-state eccentricity distribution of
the ensemble is 
\ba
n_e(e\in|e_{\rm crit})\propto\sum_{i=1}^N N_{\rm mig}^i(e_{\rm crit})
\times n_e^i(e\in),
\label{eq:ne_ss}
\ea
which can be normalized so 
$\int de\in n_e(e\in|e_{\rm crit})=1$.

\subsection{Numerical experiments}

In Figure \ref{fig:ss_e} we evolve
2000 triple systems with parameters
   $a\out/a\in=10$,  $e\out=0.5$,
  $m_s=1M_\odot$, $m\in=1M_J$,
   and $m\out=0.1M_\odot$. 
   The initial conditions are drawn 
   from uniform distributions in
  the difference in arguments of pericenter $\omega\in-\omega\out$ and
  longitudes of node $\Omega\in-\Omega\out$,   $e\in$,
  and $i_{\rm tot}$. We evolve the systems for 1000 Kozai--Lidov 
  timescales $\tau_{\rm KL}$
  (Eq.\ \ref{eq:tau_KL}) using the secular code described 
  by \citet{petro15a}, 
after ignoring all the non-Keplerian interactions
  and treating the bodies as point masses 
  (no tidal disruptions or collisions).

  For this simulation we assume that
$e_{\rm crit}=0.99$, i.e., planets that achieve maximum eccentricities
$e_{\rm in,max}>0.99$ come close enough to the host star to suffer
significant tidal dissipation and therefore migrate. In panel a, we
show the initial mutual inclination $i_{\rm tot}$ and planet
eccentricity $e\in$ for each system. Planets with $e_{\rm
  in,max}>0.99$ are shown as blue or red filled circles.  The blue
circles indicate the systems with $N_{\rm mig}(e_{\rm crit})$ (Eq.\
\ref{eq:Nmig}) larger than the median value (slower migration), while
the red circles indicate those with $N_{\rm mig}(e_{\rm crit})$
smaller than the median (faster migration).

We see that migrating systems mostly come either from regions where
$i_{\rm tot}\sim50^\circ$--$130^\circ$ and $e\in\lesssim0.6$ or from
regions with large eccentricities, $e\in\gtrsim0.7$. These are excited
to high eccentricities through, respectively, the Kozai--Lidov mechanism
or low-inclination secular oscillations due to the octupole moment of
the perturber.  The planets that migrate faster mostly start from
either high eccentricities $e\in\gtrsim0.7$ or mutual inclinations
$i_{\rm tot}\sim80^\circ-100^\circ$.

In panel b of Figure \ref{fig:ss_e}, the dashed gray line shows the time-averaged
eccentricity distribution of the planets with $e_{\rm
  in,max}>0.99$. We observe that this distribution increases 
monotonically with $e\in$, slowly for $e\in\lesssim 0.6$
and faster at higher eccentricities. To obtain the steady-state
eccentricity distribution of  warm Jupiters
we must correct for the migration speed
using Equation (\ref{eq:ne_ss}), which yields the solid black
line. With this correction, the eccentricity distribution flattens significantly
  and the peak at high eccentricities is largely 
 eliminated.   This result is expected because systems that spend
  less time at high eccentricities (small pericenter distances)
  tend to migrate more slowly and thus are more likely to be observed
  as warm Jupiters at any given time. 
 These results depend only weakly on the choice
 of the critical eccentricity $e_{\rm crit}$ as long as 
 $1-e_{\rm crit}\ll1$ (or, equivalently $r_{\rm p,crit}\ll a\in$); 
 for example, the distributions are
 almost identical for $e_{\rm crit}=0.98$ (not shown).
In other words, the eccentricity distribution from Equation (\ref{eq:ne_ss})
is approximately independent of $a\in$ provided that 
$r_{\rm p,crit}\ll a\in$.
 
In summary, we find that in these experiments the steady-state 
eccentricity distribution of migrating planets is broad and 
approximately flat. 
Also, this distribution is approximately independent of the 
semimajor axis provided that migration happens 
at $1-e\in\ll1$.

\begin{figure*}
   \centering
  \includegraphics[width=14cm]{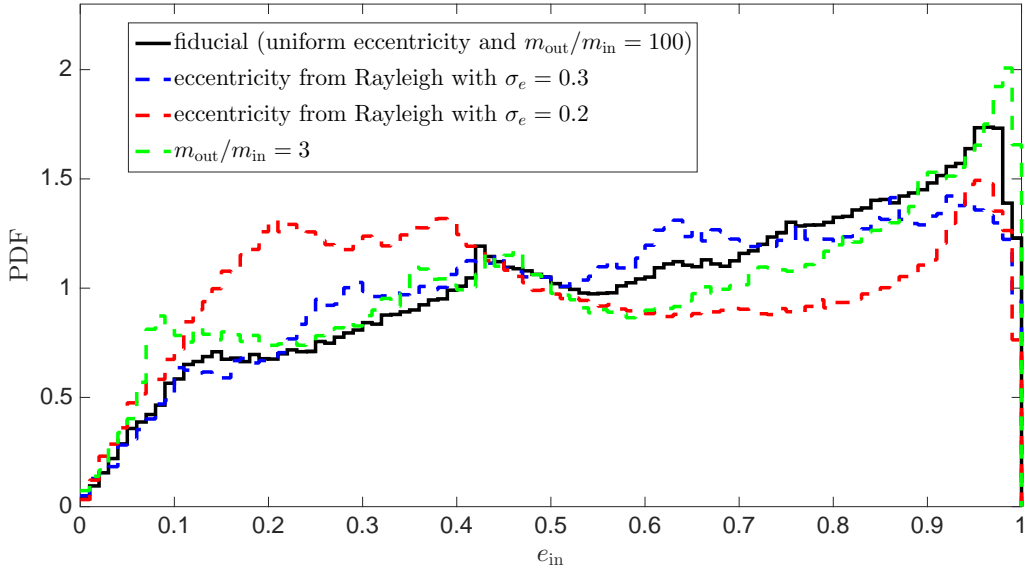}
  \caption{Steady-state eccentricity distribution
  from Equation (\ref{eq:ne_ss}), with critical eccentricity 
  $e_{\rm crit}=0.99$. The black line corresponds to our fiducial 
  simulation from Figure \ref{fig:ss_e} (same as the solid black line in
  panel b).
  The red and blue dashed lines result from initial eccentricities
  $e\in$ that follow a Rayleigh distribution 
  (Eq.\ \ref{eq:sigma_e}) with $\sigma_e=0.2$ and 0.3, 
  respectively. The dashed green line is the same as the
  fiducial simulation but with a mass ratio 
  $m\out/m\in=3$.
}
\label{fig:ss_e_param}
\end{figure*}  

\subsubsection{Effect of 
initial eccentricities and mass ratios}

\label{sec:diff}

In Figure \ref{fig:ss_e_param} we repeat the numerical experiments 
from Figure \ref{fig:ss_e} (shown as the solid black line in the
bottom panel), but starting 
with initial eccentricities of the inner orbit that 
follow a Rayleigh law, 
\ba
dp=\frac{e\,de}{\sigma_e^2} \exp(-\ffrac{1}{2} e^2/\sigma_e^2),
\label{eq:sigma_e}
\ea
where  $\sigma_e$ is an input parameter that
is related to the initial mean and rms eccentricity by $\langle
e\rangle=\sqrt{\pi/2}\sigma_e=1.253\sigma_e$ and $\langle
e^2\rangle^{1/2}=1.414\sigma_e$.
The red and blue dashed lines show the 
results for $\sigma_e=0.2$ and 0.3, respectively.

We also show the effect of changing the mass ratio of the outer
companion from a stellar-mass companion with 
$m\out/m\in=100$ ($m\out=0.1M_\odot$)
to  a planetary-mass companion with $m\out/m\in=3$ 
(green dashed line).

We observe that in all the experiments above
the expected steady-state 
eccentricity distribution for the warm
Jupiters   $n_e(e\in|e_{\rm crit})$ 
  from Equation (\ref{eq:ne_ss})  does not change 
  substantially relative to our fiducial simulation.
However, the fraction of migrating systems 
(systems with $e_{\rm max}>e_{\rm crit}$) decreases 
from $\simeq0.4$ in our fiducial
simulation to $\simeq0.26$ and $\simeq0.31$ in the simulations with
initial eccentricities drawn from a Rayleigh distribution with 
$\sigma_e=0.2$ and  $\sigma_e=0.3$, respectively, and to 
$\simeq0.28$ when the mass ratio  $m\out/m\in=3$.
This decrease in the number of planets
reaching very high eccentricities ($e_{\rm max}>0.99$)
for lower mass ratios has been observed by
\citet{teyss13}.

We conclude from these numerical experiments that the 
expected eccentricity distribution of the warm Jupiters predicted 
by our model is broad and approximately flat, with 
$\simeq40$--$50 \%$ of
  the warm Jupiters having eccentricity less than 0.5. 

\begin{figure*}[h!]
   \centering
    \includegraphics[width=17cm]{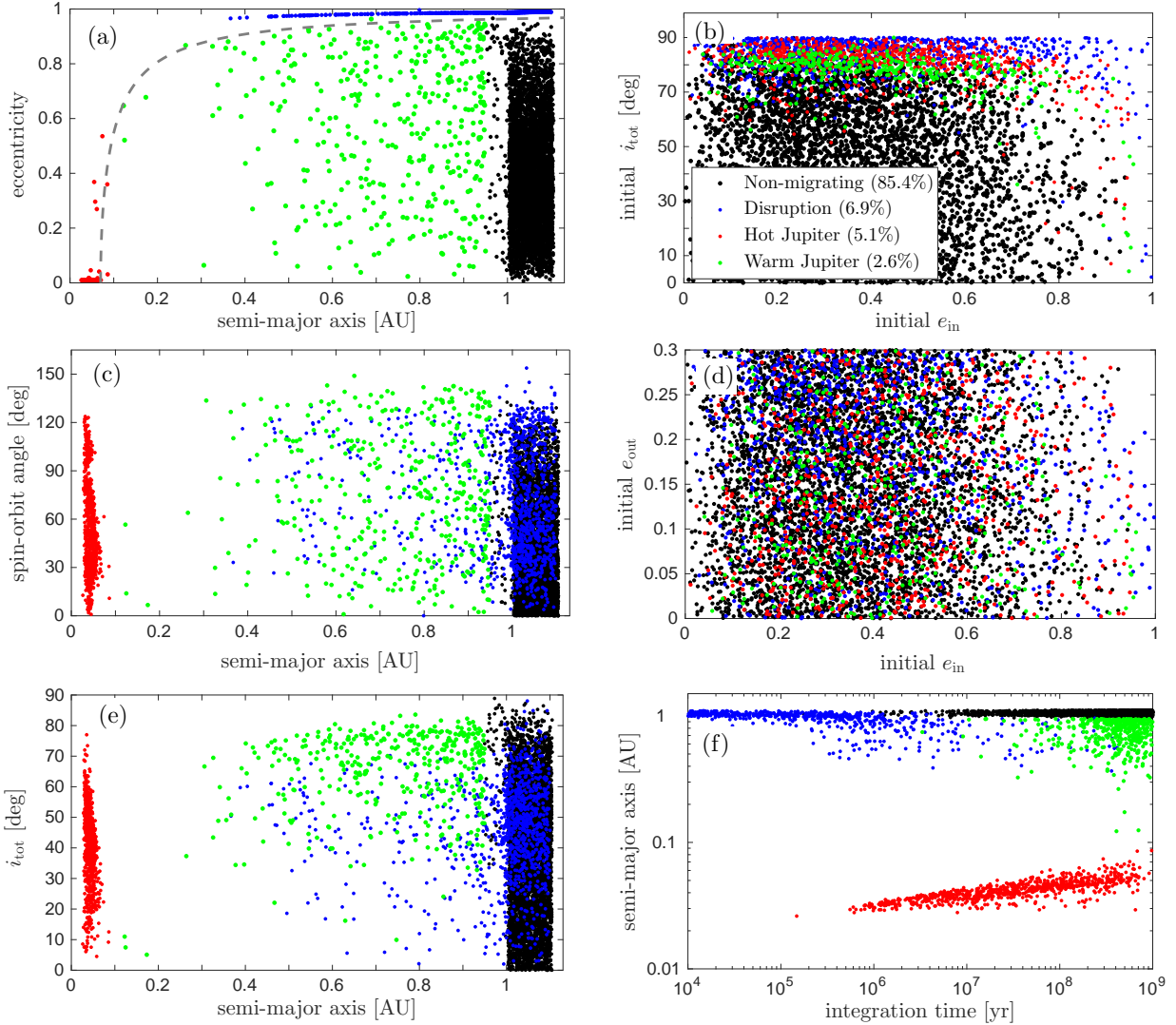}
  \caption{Outcomes for our population synthesis study, 
  as labeled in panel b.
  The planetary and stellar viscous times are 
  $t_{V,p}=0.01$ yr and $t_{V,s}=50$ yr.
The inner planet has $m\in=1M_J$ and Jupiter's radius,
and an initial semimajor axis 
drawn from a uniform distribution in $[1,1.1]$ AU. 
The outer planet
has a mass that is drawn randomly from the interval 
$[1,5]M_J$ and a semimajor axes drawn randomly from $[5,6]$ AU. 
The initial eccentricity of the inner planet is drawn from a 
Rayleigh distribution with $\sigma_e=0.3$, while the initial eccentricity 
of the outer planet and the mutual inclinations are uniformly 
distributed in the intervals $[0,0.3]$ and 
$[0,90^\circ]$, respectively. 
 {\it Panel (a):}  final eccentricity versus final semimajor axis
  of the  inner planet.
The constant angular momentum track 
$a\in(1-e\in^2)=0.07$ AU is indicated by a dashed line.
  {\it Panel (b):} the initial mutual inclination versus
  initial eccentricity of the inner planet $e\in$.
 {\it Panel (c):} final spin-orbit misalignment angle (angle between the 
 host star's spin axis and the inner planet's orbital axis ${\bf h}\in$)
 versus semimajor axis.
{\it Panel (d):}  initial eccentricity of the outer planet versus
the initial eccentricity of the inner planet.
{\it Panel (e):} final mutual inclination $i_{\rm tot}$ (angle between
${\bf h}\in$ and ${\bf h}\out$)
 versus semimajor axis.
  {\it Panel (f):}   final semimajor axis of the inner planet
  versus the time at which the simulation is stopped.}
\label{fig:pop}
\end{figure*}  

\section{Population synthesis study}
\label{sec:pop}

We ran a series of numerical experiments to study
 the evolution of triple 
systems consisting of a Sun-like host star and two orbiting
planets with masses $m\in$ and $m\out$.
We use the full set of equations of motion described in
\cite{petro15a} for hierarchical triple systems; 
they follow the orbital evolution of the inner
and outer planetary orbits and the spins of both the central
star and the inner planet including 
the effects from general relativity,
tidal and rotational bulges, and tidal dissipation.
These experiments allow us to  assess  how well 
our simple model for the steady-state eccentricity 
distribution in \S\ref{sec:ss_e}---which is based only on the 
orbit dynamics---works when extra forces
are included, and provide explicit predictions for the
properties of hot and warm Jupiters formed by 
high-eccentricity migration. 

In our experiments, 
the inner planet has Jupiter's mass and radius,
and an initial semimajor axis 
drawn from a uniform distribution in the interval $[1,1.1]$ AU,
where this narrow range in semimajor
axes is chosen for simplicity rather than realism.
The outer planet has a mass and semimajor axis that are randomly 
drawn in the intervals 
$[1,5]M_J$ and $[5,6]$ AU. 
With this choice of semimajor axes and masses 
the outer planet can in principle excite large-amplitude 
eccentricity oscillations in an inner planet with a semimajor axis as small as 
$a\in\sim 0.2$--$0.3$ AU
 without this excitation being quenched by general relativity 
 (see Eq.\ \ref{eq:acrit}).

The initial eccentricity of the inner planet
is drawn from a Rayleigh distribution 
(Eq.\ \ref{eq:sigma_e})
with   $\sigma_e=0.3$. The eccentricity of the outer planet 
$e\out$ is drawn uniformly from the interval $[0,0.3]$.
The mutual inclination of the inner and
outer planets $i_{\rm tot}$ is uniformly distributed in the interval
$[0,90^\circ]$. 

We discard systems that do not satisfy the stability condition
\citep{petro15c}:
\ba
\frac{a\out(1-e\out)}
{a\in(1+e\in)}>2.4
\left[\max(\mu\in,\mu\out)\right]^{1/3}
\left(\frac{a\out}
{a\in}\right)^{1/2}\!\!+1.15\nonumber\\
\label{eq:stability}
\ea
where $\mu\in=m\in/m_s$
and $\mu\out=m\out/m_s$. 
The systems that do satisfy this stability criterion are 
expected to evolve secularly (no exchange of orbital
energy between the planets).

The arguments of pericenter and 
longitudes of the ascending node are chosen
randomly  for the inner and outer planetary orbits. 
The host star and the inner planet initially spin with 
periods of 10 days and 10 hours, respectively. Both spin vectors are
parallel to the initial orbital angular momentum of the inner 
planet (${\bf \hat{h}}_{\rm{in},0}$), 
implying that the initial stellar and
planetary obliquities are zero relative to the orbit of the inner
planet.
We do not include spin-down due to stellar 
winds in modeling the evolution of the spin of the host star.

We stop each run when one of the following outcomes is achieved:
(i) the inner planet evolves into a hot Jupiter in a nearly circular orbit 
($a\in<0.1$ AU, $e_{\rm in }<0.01$);
(2) the inner planet is tidally disrupted, which we define to occur
when the pericenter distance 
is less than 0.0127 AU \citep{GRL11};
(3) the planet has survived for a maximum time chosen uniformly random
in the interval $[0,t_{\rm max}]$. The maximum time $t_{\rm max}$ is
chosen to be 1 Gyr. This is shorter than the typical age of the host
stars of warm Jupiters, but our results should be insensitive to
$t_{\rm max}$ so long as it is much larger than the Kozai--Lidov
and migration timescales. 
The typical Kozai--Lidov timescale is $\tau_{\rm\tiny KL}\sim5\times10^{3-4}$ yr 
(Eq.\ \ref{eq:tau_KL}). The migration timescale is determined by
the planetary viscous time $t_{V,p}$; we choose this to be 0.01 yr, to allow
planets to migrate up to $\simeq0.07$ AU 
with zero eccentricity within 1 Gyr. This is roughly
equivalent to setting $t_{\rm max}=10$ Gyr and choosing a viscous time
of $t_{V,p}=0.1$ yr (slower migration), but carrying out such
simulations would be  much more expensive.

We have compared the three-body dynamics predicted by the secular code
used in this section to direct N-body integrations using the 
high-order integrator IAS15 \citep{RS15}, 
which is part of the REBOUND package \citep{RL12}.
We carried out this comparison for a few representative cases in which
the three bodies were treated as point masses, with initial conditions
$a\out /a\in=5$, $e\out=0.3$, $i_{\rm
  tot}=\{75^\circ,80^\circ,85^\circ\}$, and $e\in=0.1$.
We found that the two codes produce similar eccentricity distributions 
of the inner planet
averaged over $\sim20$ Kozai--Lidov cycles. 
The two codes disagree on the relative phases of the 
eccentricity oscillations of the inner and outer planets, but 
this disagreement should not affect the general results 
of our population synthesis study as the planet migration depends mainly
on the eccentricity distribution after many oscillation cycles.

\subsection{Outcomes}
\label{sec:outcome}

In Figure \ref{fig:pop}, we show the initial and final 
orbital elements from our population
synthesis study, which followed 15,000 systems
that are stable according to Equation (\ref{eq:stability}).

We classify the outcomes as follows (ordered
in decreasing frequency):
\begin{enumerate}

\item {\it non-migrating (85.4\%)}: The inner planet
does not reach eccentricities 
that are high enough to induce migration. More precisely, these are
systems in which the final semimajor axis 
$a\in>0.95$ AU, compared to an initial semimajor axis in the range
$[1,1.1]$ AU  (black dots in Figure \ref{fig:pop}).
Most of these ($\sim90\%$) have mutual inclinations
$i_{\rm tot}\lesssim 70^\circ$ (panel b). The mean eccentricity 
of the planets in this category 
increases only from 0.36 to 0.41 from the initial to the final
states. 

\item {\it disruptions (6.9\%)}: The inner planet is tidally disrupted
(blue dots in Figure \ref{fig:pop}). Disruption is defined to occur
when
$a\in(1-e\in)<0.0127$ AU at some 
point of the simulation \citep{GRL11}. 
Most disruptions ($\simeq87\%$) happen very early in the 
simulation, within $1$ Myr of the start (panel f), and most of these 
systems start with high mutual inclinations (median $i_{\rm tot}\simeq86^\circ$, panel b).

\item {\it hot Jupiters (5.1\%)}: The inner planet becomes a hot Jupiter
(red dots), with $a\in<0.1$ AU.
The hot Jupiters are formed at a wide range of times between 1 Myr and
1 Gyr (panel f); the median formation time is $\sim40$ Myr.
Most of these systems start with
high mutual inclinations (median $i_{\rm tot}\simeq82^\circ$, panel b).

\item {\it warm Jupiters (2.6\%)}: The inner planet 
has semimajor axis in the range from 0.1 AU to 0.95 AU
at the end of the simulation
(green dots).
The warm Jupiters have a median integration time of 
$\sim400$ Myr (panel f) because they are followed for a  time chosen uniformly
 random between 0 and 1 Gyr (see discussion at the end of the
 preceding subsection). 
Most of these systems start with
high mutual inclinations (median $i_{\rm tot}\simeq78^\circ$, panel b).

\end{enumerate}

In our model all warm Jupiters would eventually evolve into hot
Jupiters given enough time. 
Therefore, the relative abundance between hot and warm Jupiters
is age-dependent and we expect to observe more hot Jupiters
per warm Jupiter in older systems. 

In summary, our population synthesis model predicts that in most
triple systems with parameters similar to those we have chosen 
the inner planet does not migrate significantly and its eccentricity
increases only slightly relative to its initial value.
The systems that migrate to form either hot Jupiters or warm 
Jupiters generally start  
from high mutual inclinations. The ratio of hot Jupiters to warm 
Jupiters increases with the age of the system.

\subsection{Production rate of hot and 
warm Jupiters}
\label{sec:rate}

In the study presented in the previous subsection, the fractions of
systems that form hot and warm Jupiters
are $\simeq5.1\%$ and $\simeq2.6\%$, respectively.
These fractions depend on the initial conditions,
most critically on the distribution of mutual inclinations 
(see panel b in Figure \ref{fig:pop}).
Since we only have weak observational constraints on the relevant
properties of hierarchical planetary systems, we cannot provide 
meaningful  
estimates of the production rates of
warm and hot Jupiters by this process. 
However, the ratio of 
hot to warm Jupiters seen in the simulations, $\simeq1.9$, 
is not strongly dependent on the initial conditions.
In particular, changing the 
Rayleigh distribution describing the initial eccentricity distribution
from $\sigma_e=0.3$ to $\sigma_e=0.2$ results in a similar
ratio, $\simeq2$; 
changing to a uniform initial eccentricity distribution in the range 
$[0,1]$ changes the ratio only to $\simeq2.2$. 
Similarly, changing the initial distribution 
of mutual inclinations from a uniform distribution in $[0,90^\circ]$
to a Rayleigh distribution with $\sigma_i=0.5$ ($\simeq28.6^\circ$)
changed the ratio of hot to warm Jupiters only to $\simeq2.3$. 

We can compare this ratio to the ratio of the number of systems 
with hot and warm Jupiters
in the RV sample, $40/96\simeq0.42$.
Thus, the population synthesis studies indicate that our
mechanism is more efficient at forming hot Jupiters relative to warm 
Jupiters than what is observed in the RV sample by a factor $\sim4$--$5$.

As discussed in \S\ref{sec:outcome}, the ratio of hot to warm Jupiters
is expected to increase with the age of the system.
Although we ran our simulations only for
1 Gyr, we argued in \S\ref{sec:pop} that the results from our simulations should 
be unchanged 
if we increase the maximum integration time
and the migration timescales (defined by the planet's viscous time
$t_{V,p}$) by the same factor. Thus we believe that the ratio of hot
to warm Jupiters given by our simulations is realistic. 
Note that the integration and migration timescales should also be 
constrained to reproduce (if possible) the semimajor
axis distribution of the hot Jupiters (see \S\ref{sec:sma}).

In conclusion, even though the production rates
of hot and warm Jupiters depend strongly on
the initial conditions, their relative rates do not. 
We find that the ratio between the number of systems with hot 
Jupiters and the number of systems with warm Jupiters
is roughly 2. This ratio is larger than that in the observations
by a factor of $\sim4$--5, so we expect that our mechanism
can only form up to $\sim20$--$25\%$ of the warm Jupiters even
  if it produces all of the hot Jupiters.

\subsection{Orbital distributions of hot and
warm Jupiters}

In this section we describe the distributions of orbital elements
arising from our  population synthesis study (see Figure \ref{fig:pop}).
We compare the results from our model with the observed semimajor
axis and obliquity distributions in \S\ref{sec:sma} and
\S\ref{sec:psi} respectively, while 
reserving the comparison with the eccentricity distribution for a separate
and more in-depth section in \S\ref{sec:ecc_obs}.
 
 \begin{figure}[t!]
   \centering
  \includegraphics[width=8.6cm]{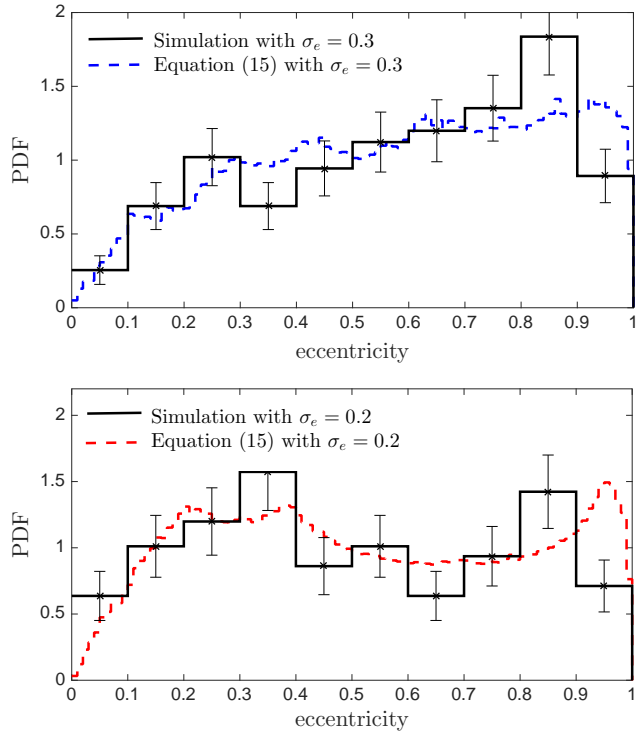}
  \caption{Eccentricity distribution of the warm Jupiters 
  from simulations (solid black line) in which the initial
  eccentricities are chosen from a Rayleigh distribution
  (Eq.\ \ref{eq:sigma_e}) with
  $\sigma_e=0.3$ (upper panel; more results from this simulation
  are shown in Figure \ref{fig:pop}). 
  The dashed blue line shows the steady-state  
  eccentricity distribution predicted from Equation (\ref{eq:ne_ss}) 
  with $e_{\rm crit}=0.99$ (same as dashed blue 
   line in Figure \ref{fig:ss_e_param}).
The lower panel shows similar
  results for $\sigma_e=0.2$.
    The error bars indicate 
  the  1$\sigma$ confidence limits from the 
  Poisson counting errors for each bin. \\}
\label{fig:e_sim_ss}
\end{figure}  

\subsubsection{Eccentricities}

In the upper panel of Figure \ref{fig:e_sim_ss} we show the eccentricity 
distribution of the warm Jupiters from the simulation shown in Figure
\ref{fig:pop}, in which the initial eccentricities are chosen from a
Rayleigh distribution with $\sigma_e=0.3$. The lower panel is similar,
but the initial eccentricities are chosen from a distribution with
$\sigma_e=0.2$. We compare our results with the
steady-state eccentricity distribution from Equation 
(\ref{eq:ne_ss})  with critical eccentricity 
  $e_{\rm crit}=0.99$ and a
  fixed perturber ($m\out=0.1M_\odot$, $a\out=10$ AU, and
  $e\out=0.5$), shown as dashed blue and red lines.

We observe that the simple model in
\S\ref{sec:ss_e} broadly reproduces the distribution 
from the simulations. However, there are small but significant
differences at very high eccentricities: 
the simulation has fewer planets at
$e>0.9$ than the model predicts, and more in the 
region $[0.8,0.9]$.
 These differences probably arise because the warm Jupiters in our 
simulations are constrained to a maximum eccentricity
given by $a\in(1-e\in^2)\gtrsim0.07$ AU
(dashed line in panel a of Figure \ref{fig:e_sim_ss})---planets
having less angular momentum than this track are either
tidally disrupted or form hot Jupiters.
In contrast, our simple model ignores tidal disruption
and circularization, allowing
for warm Jupiters with 
arbitrarily large eccentricities and thereby overpopulating the bin
$e>0.9$ relative to our simulations.

In summary, our simple model for the steady-state 
eccentricity distribution in \S\ref{sec:ss_e} 
 reproduces well the results from our population 
 synthesis study, even though it ignores the effects from general relativity,
 tides, and disruptions. The simple model predicts a somewhat larger number 
of warm Jupiters with the highest eccentricities because it does not
account for tidal disruption and tidal dissipation.

 \begin{figure}[t!]
   \centering
  \includegraphics[width=8.6cm]{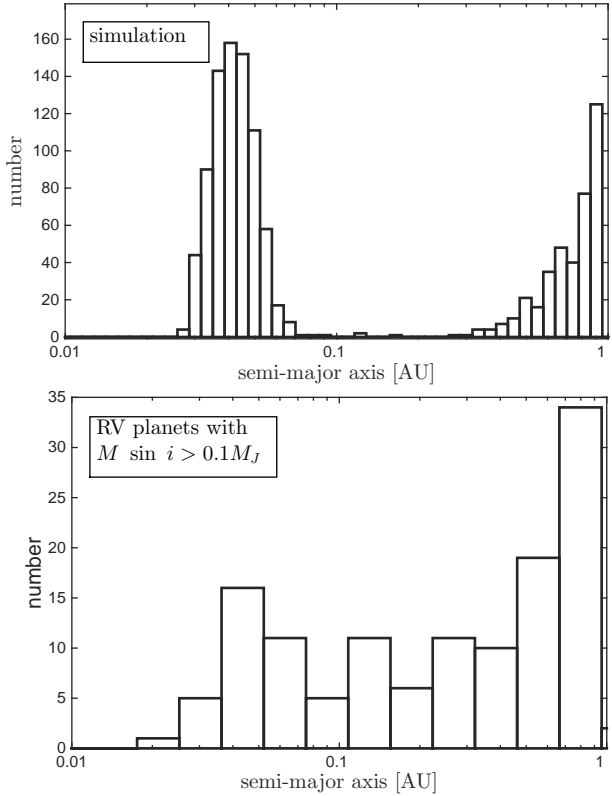}
  \caption{Semimajor axis distribution of the planets within
  $1$ AU in the population synthesis study 
  (upper panel), and in the sample of planets with $M\sin i>0.1M_J$ 
found by radial-velocity (RV) surveys
  (lower panel).\\}
\label{fig:a_sim_RV}
\end{figure}

\subsubsection{Semimajor axes}
\label{sec:sma}

In the upper panel of Figure \ref{fig:a_sim_RV} we 
show the semimajor axis distribution of the migrating
planets (hot and warm Jupiters) from the population 
synthesis study shown in Figure \ref{fig:pop}.
The lower panel shows the semimajor axis 
distribution of planets with $a<1$ AU discovered in 
radial-velocity (RV) surveys.

Our simulations
 predict that the number density of warm Jupiters per unit of $\log a$
 increases with $a$, 
 with $\simeq95\%$ of the warm Jupiters between 0.1
   and 1 AU having semimajor axes larger than 0.4 AU. 
   In contrast, the observed sample has $\simeq38\%$ of the planets in the 
   semimajor axis range inside 0.4 AU.

Our model also produces a population of hot Jupiters  at $a<0.1$ AU
with a pile-up at $\sim0.04$--$0.05$ AU.
Quantitatively,  the distribution of semimajor axes of the 
hot Jupiters in the simulation matches that of the observed hot Jupiters
 with $a\lesssim0.07$ AU in the RV sample
 ($p$-value of $\sim0.2$ from a Kolmogorov--Smirnov [KS] test).
The observed population of hot Jupiters at $a\gtrsim0.07$ AU, which
are not present in our simulation, could be 
produced by enhancing the tidal dissipation in the planet.

In summary, our population synthesis study fails to reproduce
 the observed semimajor axis distribution of warm Jupiters in the range
 $\sim0.1$--$0.4$ AU. The simulation does reproduce the shape
  of the semimajor axis distribution in the ranges 0.5--1 AU and
 inside 0.07 AU.

 \begin{figure}[t!]
   \centering
  \includegraphics[width=8.6cm]{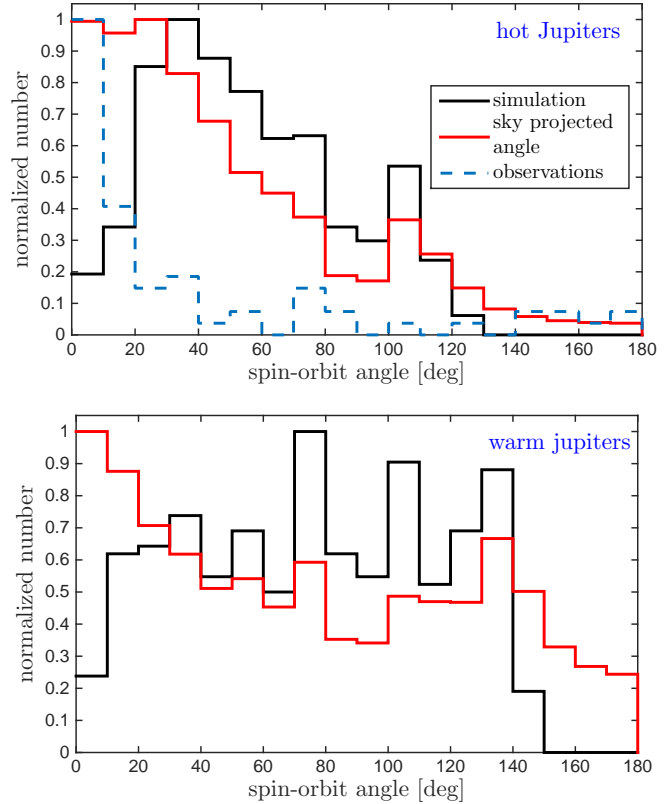}
  \caption{Final spin-orbit misalignment 
  angle (angle between the host 
 star's spin axis and the angular momentum 
 of the inner orbit ${\bf h}\in$) from our simulations (black
 lines). The upper panel shows hot Jupiters ($a<0.1$ AU) and the
 lower panel shows warm Jupiters ($0.1\mbox{\,AU}<a<1\mbox{\,AU}$). 
 The spin-orbit angle projected on the plane of 
 the sky is shown as red lines.
 The observed distribution from 65 hot Jupiters
 is shown as the blue dashed line.
\\}
\label{fig:psi}
\end{figure}  

\subsubsection{Stellar obliquities}
\label{sec:psi}

In Figure \ref{fig:psi} we show the final spin-orbit
angle (or stellar obliquity, angle between the host 
 star's spin axis and planet's orbital axis ${\bf h}\in$) of the hot and warm 
 Jupiters in our simulations (upper and
lower panels, respectively). 
We also calculate the distribution of sky-projected spin-orbit
angles (red lines) by randomizing 
the orbital configurations relative to a fixed observer. The
sky-projected angles can be directly compared to the observed 
angles as determined by the Rossiter-McLaughin effect 
(e.g., \citealt{FW09,CB14}).
For comparison we show the sample of 65 hot Jupiters 
with stellar obliquity measurements 
in the upper panel (blue dashed line). 
We do not plot the observations of warm Jupiters because the
sample size is too small for a meaningful comparison. 

We observe that the obliquities of the hot Jupiters in the simulation
are concentrated in the interval $\sim20^\circ$--$80^\circ$. The
simulations produce $\sim16\%$ retrograde hot Jupiters, similar to the
fraction in the observed sample of $9/65\simeq14\%$.  The projected
obliquity distribution of the simulated hot Jupiters (red line in the
upper panel) peaks at small values, similar to the observed
distribution.  However, the observed sample has a strong peak at low
obliquities---$38/65\simeq58\%$ of the hot Jupiters have projected
obliquities of $\lesssim20^\circ$---compared to only $\simeq27\%$ in
our simulations.  Apart from this discrepancy at low obliquities, the
observed and model distributions match reasonably well.

In the lower panel of Figure \ref{fig:psi} we show the 
obliquity distribution of the warm Jupiters, which is 
significantly broader than that of the hot Jupiters. 
In particular, $\sim40\%$ of the warm Jupiters have retrograde
obliquities compared to $\sim16\%$ of the
hot Jupiters. 
There are only 2 warm Jupiters with measured
 obliquities, too few to plot on the lower panel of Figure \ref{fig:psi}.

We note that the KL timescales in our simulated systems
are small ($\tau_{\rm\tiny KL}\sim5\times10^{3-4}$ yr 
from Eq.\ [\ref{eq:tau_KL}]) compared to the 
spin precession timescale due to the
rotation-induced stellar quadrupole ($>5\times10^5$ yr).
Therefore, the stellar obliquity distribution is unaffected 
by the secular resonance that occurs when the stellar precession 
rate matches the orbital precession rate as described in
\citet{storch14} and \cite{storch15}.

 \begin{figure}[t!]
   \centering
  \includegraphics[width=8.4cm]{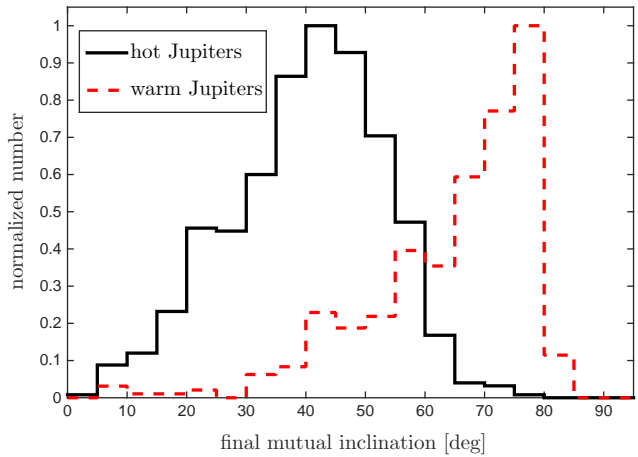}
  \caption{Final mutual inclination of
  the planetary orbits 
(angle between  ${\bf h}\in$
  and ${\bf h}\out$) for the systems with hot Jupiters
 (solid black line) and warm Jupiters (dashed red line).
\\}
\label{fig:inc}
\end{figure}  

\subsubsection{Mutual inclinations}
\label{sec:itot}

In Figure \ref{fig:inc} we show the distribution of the mutual 
inclinations between the inner and outer planetary orbits for 
the systems with hot Jupiters
(solid black line) and warm Jupiters (dashed red line).

Systems with hot
Jupiters tend to have mutual inclinations 
clustered around $i_{\rm tot}\sim40^\circ$. This peak in the
inclination distribution is a known feature of Kozai--Lidov migration in the quadrupole
approximation \citep{FT07}, which can be explained as follows. 
Let us treat the planet as a test particle for simplicity\footnote{The argument 
does not depend critically on this approximation (see \citealt{FT07}).}. 
The conservation of 
angular momentum normal to the outer orbit and the energy in 
Equation (\ref{eq:conserved}) implies that the following is a 
conserved quantity:
\ba
\mathcal{K}=e\in^2\left( 5\sin^2\omega\in\sin^2i\in-2 \right).
\label{eq:K1}
\ea
A planet can migrate to form a hot Jupiter when it reaches a
sufficiently large maximum eccentricity $e_{\rm max}$. At this point
the inclination is a minimum $i_{\rm min}$ (or a maximum if the orbit
is retrograde) because of the 
conservation of $[1-e\in^2]\cos^2 i_{\rm in}$. The maximum
eccentricity in a Kozai--Lidov cycle is achieved 
at  $\sin^2\omega\in=1$. Thus, 
\ba
\mathcal{K}=e_{\rm max}^2\left( 5\sin^2i_{\rm min}-2 \right).
\label{eq:K2}
\ea
If the orbit passes through $e\in=0$ at any point in the cycle, Equation (\ref{eq:K1}) 
implies that $\mathcal{K}=0$ and from Equation (\ref{eq:K2}) we
have $\sin^2 i_{\rm min}=\frac{2}{5}$ ($i_{\rm min}=39.2^\circ$ and 
$140.7^\circ$). 
Therefore, if we start from nearly circular orbits, we expect 
that the hot Jupiters should have a distribution of mutual inclinations 
that peaks strongly at $i_{\rm tot}=39.2^\circ$ for prograde orbits. 

The dispersion around $i_{\rm tot}\simeq40^\circ$ in our simulations is
$\simeq13^\circ$. This dispersion has several distinct causes: 
(i) the initial eccentricity of the inner planet is not precisely zero,
as it follows a Rayleigh distribution with $\sigma_e=0.3$;
(ii) the octupole-level forcing from the outer 
companion allows for migration from orbits with mutual inclination
lower than  $\sim40^\circ$ \citep{petro15b};
(iii) the quantity $(1-e\in^2)\cos^2 i_{\rm in}$
is not precisely conserved when the octupole-level perturbations from
the outer companion are included \citep{NF14}.

The peak of the distribution of mutual inclinations is relatively
insensitive to the initial eccentricity distribution: it 
 changes from $i_{\rm tot}\simeq40^\circ\pm 13^\circ$ 
for an initial eccentricity distribution drawn 
from a Rayleigh distribution with $\sigma_e=0.3$
to  $i_{\rm tot}\simeq41^\circ\pm 11^\circ$ 
and $i_{\rm tot}\simeq38^\circ\pm 16^\circ$ 
for a Rayleigh distribution with $\sigma_e=0.3$
and uniform in $[0,1]$, respectively.
Also, we note that our initial conditions are constrained to 
$i_{\rm tot}\leq90^\circ$ and if we relax this assumption
allowing for $i_{\rm tot}=0-180^\circ$ we would expect 
an extra peak at $\sim140^\circ$ \citep{FT07}.

The warm Jupiters generally have much larger mutual
inclinations than the hot Jupiters (Fig.\ \ref{fig:inc}). 
This difference arises mainly because 
the mutual inclination of the warm Jupiters reflects
the time-averaged distribution of the ensemble,
not the minimum values as it is case for the hot Jupiters.

In summary, the systems with hot Jupiters
in our population synthesis study tend to
have outer planetary companions with 
mutual inclinations  clustered at around $\sim40^\circ$, 
while most of the systems with warm Jupiters have 
companions with mutual inclinations 
$\gtrsim60^\circ$.

\begin{figure*}
   \centering
  \includegraphics[width=14cm]{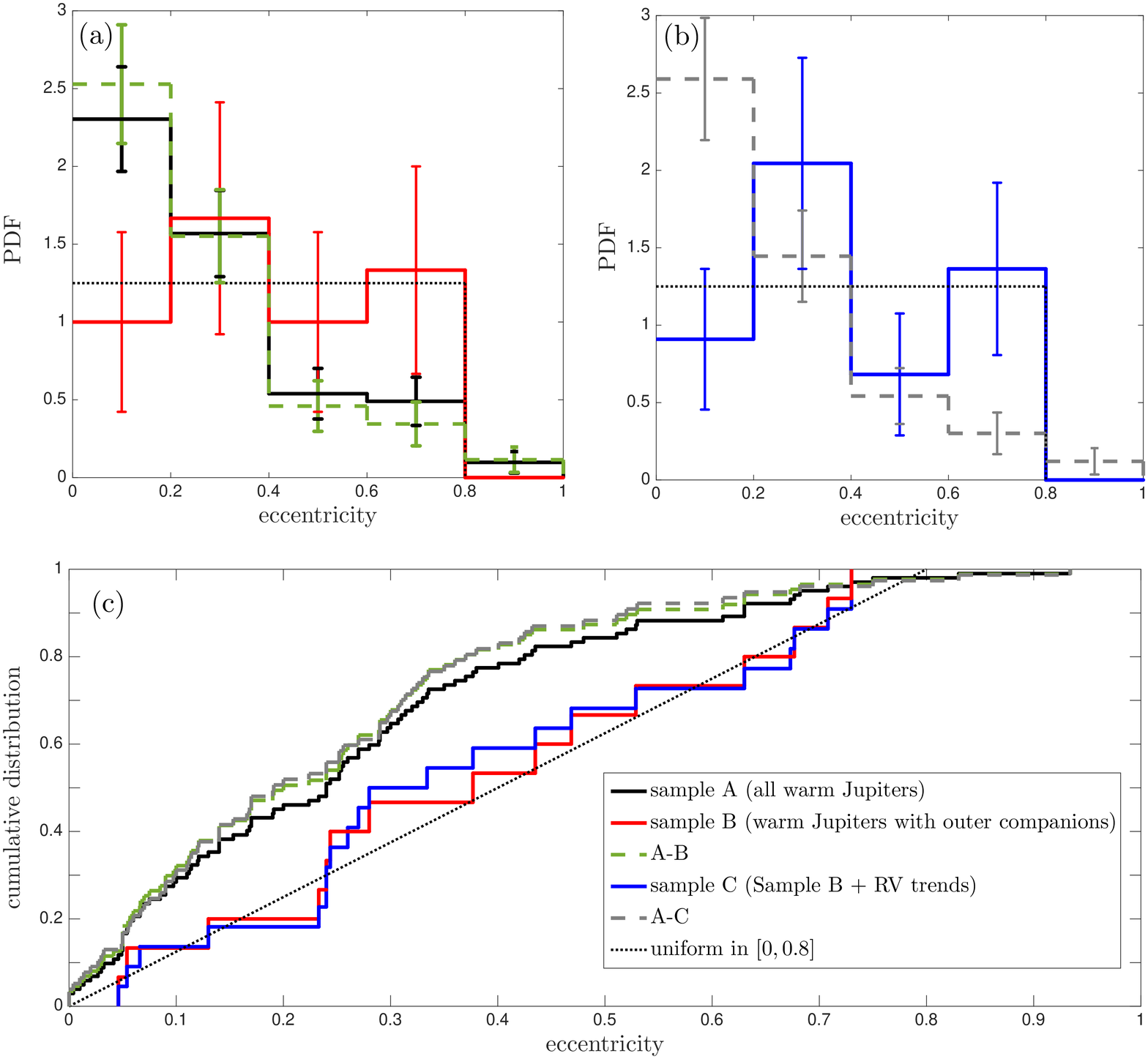}
  \caption{
Density (panels a and b) and cumulative (panel c)
  distributions of the eccentricities of the warm Jupiters discovered
  in RV surveys.   The solid black lines show the whole sample of 
  warm Jupiters (sample A; 102 planets in 96 systems).
  The solid red lines show the sample of warm Jupiters
  with outer planets at  $a\out>1$ AU
  and no companions within 1 AU
  (sample B; 15 planets in 15 systems).
  The solid blue lines show the sample B 
  augmented  by systems containing warm Jupiters with $a>0.3$ AU 
  and linear RV trends indicating a distant companion (sample C, 22 planets in 22 systems).
The dashed green (gray) lines show the samples A$-$B (A$-$C), while the
dotted black line shows a uniform distribution in 
  $[0,0.8]$ for reference. The error bars in panels a and b
  indicate the 1--$\sigma$ Poisson errors for each bin.
  }
\label{fig:hist}
\end{figure*}

\section{Eccentricity distribution of the 
observed warm Jupiters}
\label{sec:ecc_obs}

We compare the eccentricity distribution in our simulations to
three samples of exoplanets\footnote{We take the data from http://exoplanets.org
    \citep{wright11} and http://exoplanet.hanno-rein.de/
    \citep{rein12} as of September 2015.}:

\begin{description}

\item[A] All exoplanets discovered in radial-velocity surveys with
  masses $M\sin i>0.1M_J$ and semimajor axes $a=0.1$--1 AU (these
  limits correspond to the definition of warm Jupiters used throughout
  this paper).  
  This sample consists of 102 planets in 96 planetary
  systems, and has a mean eccentricity of 0.26. In Figure
  \ref{fig:hist} we show the density (panel a) and cumulative
  distribution (panel c) of the eccentricities in 
  this sample (solid black lines). The density distribution peaks at
  $e\lesssim0.2$ and decays monotonically for higher eccentricities.
  In contrast, our model predicts a flat eccentricity distribution
  (no peak at $e\lesssim0.2$) and thus does not match the 
  observed distribution in this sample. 

\item[B] The planets in sample A that have relatively well-separated
  outer companion planets ($a\out\ge 1$ AU and
  $a\out/a\in>2$\,\footnote{This choice is somewhat arbitrary. It
    excludes two systems with planets in 2\,:\,1 mean-motion
    resonances, HD 82943 \citep{tan13} and HD 73526 \citep{tinney06},
    whose orbital configuration are likely due to disk
    migration.}. The motivation for this choice is that our model
  predicts that the warm Jupiters should have planetary companions in
  orbits outside $\sim1 $ AU.  This sample consists of 15 planets; the
  mean eccentricity of 0.38 is significantly larger than in sample
  A. We show the eccentricity distribution of this sample in Figure
  \ref{fig:hist} using solid red lines. The distribution is flatter
  than sample A; a KS test between the distribution in the samples
  A$-$B (dashed green lines) and B results in a p-value of 0.08.
  Sample B is consistent with a uniform distribution in the
  eccentricity range $[0,0.8]$ (dotted black lines in Figure
  \ref{fig:hist}).

\item[C] Sample B augmented by systems from sample A that exhibit
  linear trends in radial velocity, which indicate the presence of a
  long-period planet or a stellar companion.  These companions are
  likely to be far enough so the eccentricity oscillations of the warm
  Jupiters with the smallest semimajor axes are quenched by
  relativistic precession (see discussion in \S\ref{sec:conditions}).
  Thus, we only add a system with a radial-velocity trend if the
  semimajor axis of the warm Jupiter is $a>0.3$ AU (see Eq.\
  \ref{eq:acrit}).  This sample totals 22 planetary systems and is
  shown by solid blue lines in Figure \ref{fig:hist}. This
  distribution is similar to the distribution of sample B and
  significantly flatter than of sample A; a KS test
  between the distribution in the samples A$-$C (dashed grey lines)
  and C results in a p-value of 0.02.

\end{description}

The finding that warm Jupiters with outer planetary 
companions have a flatter
eccentricity distribution than the whole sample
of warm Jupiters is originally due to 
\citet{DKS14}. We confirm their results, but for a 
bigger sample: 9 planets in \citet{DKS14}
compared to 15 planets (or 22 considering the RV trends) in our
study. The difference in the sample size is due to
the larger ranges of masses we consider
($M\sin i>0.1M_J$ compared to 
$M\sin i>0.3M_J$ in \citealt{DKS14}) and the larger range of 
 semimajor axes 
 ($a<1$ AU in our work compared to 
$a<0.5$ AU in \citealt{DKS14}). 

In summary, we observe that the eccentricity distribution of
the sample of warm Jupiters is skewed towards low
eccentricities, but the subsample of systems with 
outer planets (either RV detections or RV linear trends)
is approximately flat for eccentricities in the range $[0,0.8]$.

\begin{figure}
   \centering
  \includegraphics[width=8.5cm]{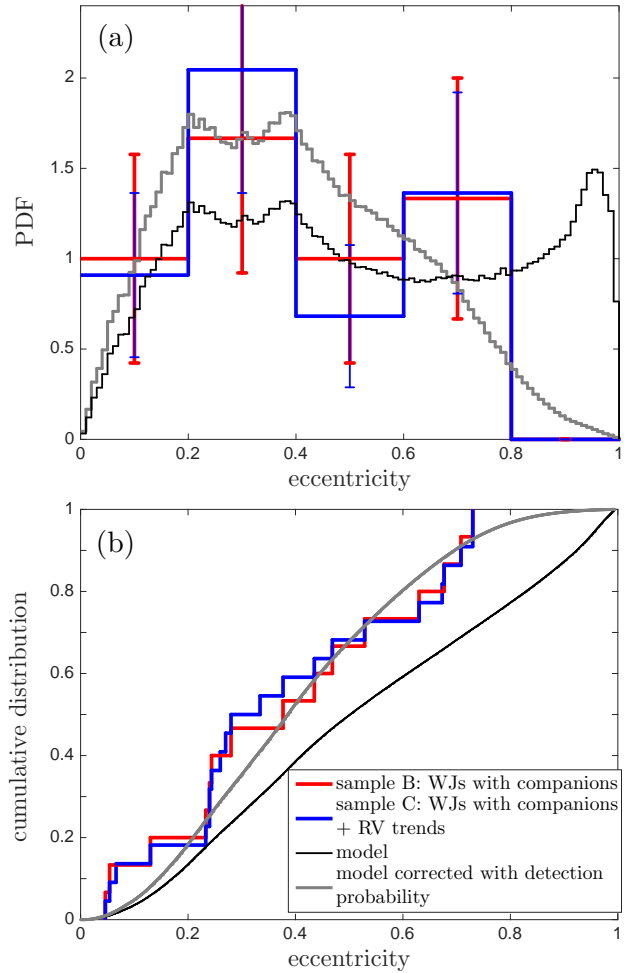}
  \caption{Density (panel a) and cumulative (panel b) 
  distributions from our simulation
  with the initial
  eccentricity distribution drawn from a Rayleigh 
  distribution with $\sigma_e=0.2$ (solid black line,
same as the dashed red line in Figure 
  \ref{fig:ss_e}).
  The solid gray lines indicate the effect of correcting
  the eccentricity distribution by the factor $P_{\rm detect}(e|e_{\rm t},e_{\rm sd})$ in 
  Equation (\ref{eq:pdetect}) with $e_{\rm t}=0.75$ and
$e_{\rm sd}=0.15$. 
  The observed eccentricity distributions in samples B and C of
  \S\ref{sec:ecc_obs} are shown as solid red and blue lines. }
\label{fig:model_obs}
\end{figure}  

\subsection{Comparison with our model}

In \S\ref{sec:ss_e} we showed that the steady-state
eccentricity distribution predicted by our model is relatively flat
in $[0,1]$ and largely independent of 
the initial distribution of eccentricities of the inner
planet (see Figure \ref{fig:ss_e_param}).

In Figure \ref{fig:model_obs}, we compare
the eccentricity distribution predicted by
our model in which the initial eccentricities are drawn
from a Rayleigh distribution with $\sigma_e=0.2$
(solid black lines) to samples B and C. 
We observe that the model reproduces, at least qualitatively,
the overall flat profile of the observed eccentricity distribution 
up to $e\sim0.8$.
However, the model fails to account for the lack of planets 
with $e\gtrsim0.8$ in the observations.

One possible reason for the absence of highly 
eccentric ($e\gtrsim0.8$) planets
in the observations is 
the selection effects in RV surveys against
detecting planets with $e\gtrsim0.6$---sparse observations of 
high-eccentricity orbits are likely to miss the strong reflex 
velocity signal near pericenter, leading to non-detection 
of planets that would be detected at the same semimajor 
axis and smaller eccentricity
\citep{cumming04,otoole09}. This possibility is discussed further 
in the following subsection.

\subsubsection{Correcting for eccentricity selection bias}
\label{sec:bias} 

We briefly discuss how observational selection biases against
detecting high-eccentricity planets \citep{cumming04} could bring 
our results into closer agreement with the observations. We emphasize
that this discussion is for illustration only and is not a quantitative
analysis of the effects of selection bias. 

Motivated by the results in  \citet[][Figure 4]{cumming04} 
we parametrize the eccentricity dependence of 
the detection probability in radial-velocity surveys as
\ba
P_{\rm detect}(e|e_{\rm t},
e_{\rm sd})=\frac{1}{2}\left[1+{\rm erf}
\left(\frac{e_{\rm t}-e}{e_{\rm sd}}
\right)\right],
\label{eq:pdetect}
\ea
where $e_{\rm t}$ represents the detection threshold for a 
given signal-to-noise ratio and number of
measurements (for $e=e_{\rm t}$ the signal  is detected
half of the time). 
The quantity $e_{\rm sd}$ represents the characteristic
width of the detection probability around $e_{\rm t}$.

In Figure \ref{fig:model_obs} 
we show the results of the model eccentricity
distribution corrected by Equation
(\ref{eq:pdetect}) with $e_{\rm t}=0.75$ and
$e_{\rm sd}=0.15$ (solid gray lines).
Although this choice of parameters is somewhat arbitrary,
it is motivated by one of the models of \citet{cumming04}, who shows
that the detection probability  with signal-to-noise ratio 10
and $N=39$ observations is 1 for $e\lesssim0.6$ and
drops to 0.5 and 0.1 for $e\sim0.75$ and $e\sim0.9$,
respectively.

Our corrected eccentricity distribution
fits the observed distribution very well.
The $p$-values from a KS test comparing our 
model distribution with samples B and C increase from $\sim0.07$  and
$\sim0.02$
when  no correction is applied to 
0.88  and 0.44 when we correct the distribution.

These results are largely independent
of the initial eccentricity distribution in the simulations and similar results 
are obtained for a uniform distribution and a Rayleigh
distribution with $\sigma_e=0.3$.
Our simulated eccentricity distributions 
can match the data on warm Jupiters with outer
companions for various functional forms of the
detection probability provided that it drops
from $\sim1$ at $e\sim0.6$ to small values (say 
$\sim0.1$) at $e\sim0.9$.

In summary, our model can match the observed
eccentricity distribution of the warm Jupiters with outer 
companions when plausible selection biases against 
detecting highly eccentric planets are taken into account.


\section{Discussion}

We have examined the expected eccentricity distribution of warm
Jupiters that migrate through a combination of secular gravitational
interactions with a distant companion and tidal dissipation
(``high-eccentricity migration'').  Our main
result is that the expected distribution is approximately flat
($dn/de\sim\mbox{constant}$).  This result partially resolves the
well-known problem that warm Jupiters cannot have reached their
current semimajor axes through high-eccentricity migration because
their pericenter distances are too large to allow for significant
tidal dissipation by the host star.

However, our model cannot fully reproduce the
eccentricity distribution of the warm Jupiters
because it does not produce enough planets with eccentricities
$\lesssim 0.2$ (Figure \ref{fig:hist}). This discrepancy is eliminated
if we consider only warm Jupiters in systems containing outer
planetary companions or linear RV trends suggesting a distant stellar
or planetary companion---and of course such a companion is required in
our model.

The population of low-eccentricity warm Jupiters
($e\lesssim0.2$) must be largely formed by a different
mechanism. Most likely, these planets
acquired their current orbital configurations 
when the gaseous disk was still present, either by
disk migration \citep{GT80,W97} or in-situ
formation (e.g., \citealt{BBL15,boley15,HWT16}).

In what follows, we describe other results and  
predictions from high-eccentricity migration and compare our
work with previous studies.

\subsection{Contribution from our model to 
the hot Jupiter population}

We discuss how the hot Jupiter population produced in our 
simulations of high-eccentricity migration
compares to the observed population. 

First, in our simulations the ratio between the number of hot 
Jupiters and the number of gas 
giant planets is $\sim5$--$7\%$.
This ratio is consistent with the observed rate of $\sim3$--$10\%$,
derived using the occurrence rate of hot Jupiters,
$\sim0.5$--$1.5\%$ \citep{gould06,mayor11}, and that
of gas-giant planets at AU distances, $\sim15\%$
\citep{mayor11}.
However, the hot Jupiter production rate derived from our models 
depends strongly on the initial conditions because the 
planets migrate only if they initially have either high mutual
inclinations or large eccentricities 
(panel  b in Figure \ref{fig:pop}).

Second, as discussed in \S\ref{sec:sma} and shown in Figure
\ref{fig:a_sim_RV}, our model predicts that the semimajor axes of
hot Jupiters are strongly concentrated in the range $\sim0.04$--$0.05$
AU (orbital period of $\sim3$--$4$ days), roughly consistent with the
distribution observed in both the RV and {\it Kepler} samples
\citep{SMT15}.  
This pile-up arises because there is a
minimum pericenter distance at which the eccentricity excitation is
limited by extra precession forces (general relativity and/or
tidal bulges, see \citealt{WL11} and \citealt{petro15b}).

Third, the  obliquity distribution
predicted by our population synthesis model is
broad (see upper panel in Figure \ref{fig:psi}).
We successfully reproduce the observed ratio
between the numbers of retrograde and prograde systems, 
but fail to explain the sharp peak in the distribution at
projected obliquities $\lesssim 20^\circ$ produced by
$38/65\simeq58\%$ of the observed sample of hot Jupiters. 
Roughly $60\%$ of the hot Jupiters with low
obliquities ($\sim40\%$ of the current sample), 
must be produced by a different 
mechanism.
 
In summary, based on the production rates, the semimajor 
axis and stellar obliquity distributions we suggest that
high-eccentricity migration can account for most of the
hot Jupiters. 
A fraction of the low-obliquity hot Jupiters must be formed
by a different mechanism.
 
\subsection{Contribution from our model to 
the warm Jupiter population}

We discuss how the warm Jupiter population produced in our simulations
compares to the observed population.

Our model can explain the eccentricity distribution of warm Jupiters observed
in samples B or C of \S\ref{sec:ecc_obs}, i.e., 
the 15 systems with outer planetary companions or the 22
systems with companions or RV trends. For comparison the total number
of systems with warm Jupiters is 
96.  Thus, our model can potentially
account for $15/96\simeq16\%$ to $22/96\simeq23\%$
of the whole population of warm Jupiters---more in the likely case
that not all companions have been detected so far.

An independent estimate of the production rate of warm
Jupiters is given by our population synthesis models.
As discussed in \S\ref{sec:rate}, we find that the relative rates of
production of hot and warm 
Jupiters do not depend strongly on the initial conditions, and our
model predicts that the ratio of hot Jupiters to warm Jupiters should
be roughly 2. This is larger than the observed ratio in the RV sample,
$40/96\simeq0.42$, by a factor of $\sim4$--$5$.
Therefore, if our mechanism accounts for most
of the hot Jupiters (see previous subsection), then we expect 
that it also accounts for $\sim20$--$25\%$ of the
systems with warm Jupiters, consistent with the estimate from the
preceding paragraph.

In summary, by comparing the relative production
rates of hot and warm Jupiters with the inferred
values from observations, we conclude that our
mechanism can account for $\sim20$--$25\%$ of
the systems with warm Jupiters.
This number matches the fraction of $\sim20\%$ derived
from the eccentricity distribution of warm Jupiters in systems 
with outer companions.

\subsection{Predictions}

We have shown that
our model can account for most of the hot
Jupiters and $\sim20\%$ of the warm Jupiters.

Two natural predictions from the high-eccentricity migration 
model discussed in this paper are:  (i) the presence 
of long-period planetary companions in most of the hot 
Jupiter systems; (ii)  the presence 
of long-period planetary companions in at least $\sim20\%$
of the warm Jupiter systems, typically those in which the warm Jupiter
has $e\gtrsim0.2$.

Recent studies show that the occurrence rate of distant
($a=5$--$20$ AU) planetary-mass
companions of hot Jupiters is 
$75\pm 5\%$, which is consistent with the prediction that high-eccentricity 
migration can account for most of the hot Jupiters \citep{knutson14,bryan16}. 
Similarly, \citet{bryan16} constrain the occurrence rate
of long-period ($a=5-20$ AU) planetary-mass companions of 
warm Jupiters to $48\pm 9\%$. This number is larger 
than the $\sim20\%$ predicted by our model
of high-eccentricity migration, but we expect that a fraction of these systems 
with the most distant of the planetary companions
do not contribute to the warm Jupiter formation
because the KL oscillations can be quenched by GR 
(see Eq.\ \ref{eq:acrit}).


In what follows, we discuss additional observational tests
that can further constrain our model using upcoming 
observational capabilities.

\subsubsection{Mutual inclination angles of hot
and warm Jupiters
systems with outer companions}

As discussed in \S\ref{sec:itot} our model predicts 
that the systems with hot and warm Jupiters should have 
outer planetary companions in inclined orbits.

For hot Jupiters, we find that the distribution of mutual
inclinations peaks at $\sim40^\circ$ with a spread of $\sim15^\circ$
around this value (see the black line in Figure \ref{fig:inc}).  As
discussed in \S\ref{sec:itot}, this peak is a feature of the Kozai--Lidov
oscillations of the inner planet in which the planet circularizes when
it reaches a maximum eccentricity, which generally happens at a
minimum inclination close to $\sim40^\circ$.

In contrast, the warm Jupiters should have companions 
with larger mutual inclinations than the hot
Jupiters, typically in the range
$\sim60^\circ$--$80^\circ$ (see Figure \ref{fig:inc}).
As described in \S\ref{sec:itot}, the larger value arises because hot
Jupiters are found at the minimum inclination achieved during a Kozai--Lidov
cycle, while the warm Jupiters represent the steady-state inclination 
distribution during a cycle. 

At present little is known about the mutual inclinations of massive
planets. A notable exception is an astrometric measurement of the
mutual inclination of $\upsilon$ And c and d, $i=30^\circ\pm1^\circ$
\citep{mcarthur10}. We expect that the {\it GAIA} 
space telescope can measure 
this angle for many of the hot and warm Jupiters with
detected outer companions out to $\sim4$ AU
(e.g., \citealt{casertano,sozzeti}).

\subsubsection{Spin-orbit angles of warm Jupiters}

As discussed in \S\ref{sec:psi} we expect that the
warm Jupiters formed by secular planet-planet
interactions will have an approximately uniform distribution of
stellar spin-orbit angles 
in the range $\sim0$--$140^\circ$ (see the lower panel in
Figure \ref{fig:psi}). 
Future space missions such as {\it TESS} \citep{ricker14}
and  {\it PLATO} \citep{rauer14} will discover 
hundreds or even thousands
of warm Jupiters in bright stars amenable
to Rossiter-McLaughin measurements of the projected spin-orbit angle. 
Ground-based transit surveys are also expected to find 
many warm Jupiters, with periods $\gtrsim10$ days (see, e.g., 
\citealt{bakos13} for a discussion on the expected rates). 
Two recent examples are 
HATS-17b with a period of 16.3 days \citet{brahm16} and
WASP-130b with a period of 11.2 days \citealt{hellier16}.

It would be particularly informative to compare the
obliquities of the warm Jupiters in eccentric orbits with outer
companions, which can likely be explained by high-eccentricity
migration, to a sample of low-eccentricity warm Jupiters without 
outer companions, which must have formed by a different mechanism and
presumably have much smaller obliquities. 

\subsection{Comparison with other work}

We discuss our results in the context of recent work 
on high-eccentricity planet migration. 

\citet{DC14} suggested that
a population of 6 systems with 
eccentric warm Jupiters and outer planetary companions
might be undergoing Kozai--Lidov
migration. This claim is 
based on the clustering of the relative apsidal angles of 
these planets at around $\sim90^\circ$.  
Since the outer companions in these systems are 
eccentric  ($e\sim0.1$--$0.4$) and the Kozai--Lidov oscillations 
are  modulated by the octupole timescale,
the authors argue that these planets are undergoing  
a slow
version of Kozai--Lidov migration, which 
preferentially produces warm Jupiters over hot Jupiters.

Our results are consistent with \citet{DC14} in the
sense that we find that in a steady state the planets 
that are most likely to be observed as warm Jupiters 
tend to have a significant octupolar modulation 
of the eccentricity oscillations (because
such planets spend less time at very high 
eccentricities and therefore have slower migration rates).
In particular, in Figure \ref{fig:n_e_erg_quad} we show the family of 
eccentricity distributions for a migrating planet in two 
limits: the quadrupole approximation (no octupole contribution)
and the ergodic approximation (strong octupole contribution).
This figure shows that in the quadrupole approximation the eccentricity
distribution diverges as $e\to1$, while in the ergodic approximation 
the distribution is a decreasing function of 
the eccentricity as $e\to1$. 
However, contrary to the results by \citet{DC14}, we do not observe a 
significant clustering of the relative apsidal angles 
($\Delta\varpi_{\rm inv}$, the difference in the longitudes of 
pericenter measured relative to the invariable plane) 
around $\sim90^\circ$ in our simulated systems 
containing warm Jupiters. These results suggest that either 
(i) the clustering observed by Dawson \& Chiang, which looks 
persuasive but is based on only six systems containing a 
warm Jupiter and an outer companion, is an unlikely 
statistical fluke; (ii) the clustering arises through some 
unrecognized selection effect; or (iii) the clustering arises for 
particular ranges of other other orbital elements that are much 
more densely populated in real systems containing warm 
Jupiters than in our simulations. 
These issues deserve further investigation.

\citet{petro15b} proposed that most hot Jupiters,
but almost no warm Jupiters,
could be formed by a process he called ``coplanar high-eccentricity 
migration''---secular
interactions of two nearly coplanar eccentric planets.
Consistent with these results we find that
planets starting with low mutual inclinations and high
eccentricities migrate rapidly (see panel a of
Figure \ref{fig:ss_e}) and are, therefore, expected to
produce hot Jupiters rather than warm Jupiters.

\citet{FH15} have recently studied the possibility that 
a population of warm Jupiters migrating through the 
KL mechanism can be depleted as the host star evolves 
off the main sequence and grows in radius. 
These authors study the eccentricity
distribution of the warm Jupiters and, in contrast to the present paper (see Figure \ref{fig:e_sim_ss}), find that the 
eccentricity distribution is strongly peaked at low 
eccentricities (Figure 17 of Frewen \& Hansen).  
We may understand this difference from the particular set of 
initial conditions used by \citet{FH15}, in which the inner
and outer planet both start on nearly circular orbits ($e\lesssim0.1$). 
This choice limits the family of eccentricity distributions in 
Figure \ref{fig:n_e_erg_quad} to only one,
which would be approximated by the dashed black and blue lines in the
top panel of Figure
\ref{fig:n_e_erg_quad} (quadrupole approximation with 
$\psi\sim50^\circ$).  
However, their calculations do not show the
extra lower-amplitude peak at $e\in=1$ found in our quadrupole distributions,
possibly because this peak is shifted to a range of lower values in
$e\in\sim0.8-0.95$ since 
their warm Jupiters are initialized at $a\in=0.1-0.45$ AU and 
reach minimum pericenter distances $a\in(1-e\in)\gtrsim 0.02$ AU.
Note that even if the inner planet starts in a circular orbit, 
but the perturber is eccentric so its potential has a significant octupolar 
moment, the expected eccentricity 
distributions are generally not peaked at low eccentricities
(see solid lines in Figure \ref{fig:n_e_erg_quad}).

 Very recently, \citet{antonini16} carried out a numerical study 
of the formation of hot and warm Jupiters from Kozai--Lidov
oscillations induced by planetary perturbers. 
Consistent with our main result, their simulations
show that the warm Jupiters formed by this mechanism 
have a nearly flat eccentricity distribution.
However, the authors claim that 
most of the observed warm Jupiters with outer planetary companions are not 
formed through high-eccentricity migration. 
Their claim is based on the observation that if the inner planet in these systems
started migration beyond $\sim1$ AU, then 
these systems would have been dynamically unstable. While the
dynamical stability of the pre-migration system is an important
constraint on the observed systems, all of the systems
in our population synthesis study are stable (according to Equation
\ref{eq:stability}) so this result does not affect our numerical
experiments.

\subsection{Stellar binary vs. planetary-mass outer 
companion}
\label{sec:binary}

The main results of this paper are valid whether the outer perturber is 
a planet or a star (see, e.g., Figure \ref{fig:ss_e_param}).
We have assumed a planetary-mass companion in most of our discussion,
motivated by the observation that 
warm Jupiters in systems with outer planets have a broad 
eccentricity distribution consistent with our model, whereas
  those in systems without outer planets mostly have $e\lesssim 0.2$.
  Here we briefly discuss what conditions are required for 
the stellar-companion
scenario to explain the warm Jupiters.

As discussed in \S\ref{sec:conditions}, there are at least two conditions
for a migrating warm Jupiter
to be undergoing eccentricity oscillations:
\begin{itemize}

\item { \it The migration rate must be slow relative 
to the secular perturbations}.
As shown by \citet{petro15a}, the Kozai--Lidov
mechanism in wide stellar binaries ($>100$ AU)
generally leads to fast migration and, therefore, produces almost no 
warm Jupiters (similar results are shown in \citealt{ASL15}).
This behavior is mostly due to the strong octupole-level 
modulation of the Kozai--Lidov mechanism in stellar binaries, which 
pumps up the eccentricity to extremely high values, enhancing 
the fraction of the planets
that migrate rapidly (in the sense of Equation \ref{eq:r_p}) or are
tidally disrupted. 
We expect that this effect is more pronounced
in stellar binaries for two different reasons.
First, stellar binaries are more eccentric
(mean eccentricity of $\sim0.6$ at separations 
of $\sim50-200$ AU; e.g,  \citealt{TK15})
than the cold Jupiters (mean eccentricity
of $\sim0.25$) and, therefore, the octupole forcing can 
have more dramatic
effects in stellar binaries (e.g., \citealt{LN11,katz11}).
Second, contrary to the case of stellar companions, 
the orbits of planetary companions can change their angular 
momentum significantly in timescales comparable 
to that of the octupole, often leading to less extreme 
eccentricities (e.g., \citealt{teyss13}). 

\item  {\it Precession due to secular perturbations 
has to be faster
than the precession due to general relativity}.
From Equation (\ref{eq:acrit}) we observe that this requirement
implies that warm Jupiters within  $a\sim1$ AU  must have 
stellar binary companions with masses of $1M_\odot$ ($0.1M_\odot$)
within $\sim200$ AU ($\sim100$ AU).
The frequency of stellar companions at these orbital separations 
in warm-Jupiter systems remains largely unconstrained, and the
constraints on stellar companions in hot-Jupiter systems are difficult
to interpret. 
Recently \citet{Piskorz15} searched for low-mass
stellar companions within $\sim100$ AU around
systems with hot Jupiters (not warm Jupiters).
They found no evidence of an excess of binary companions
relative to field stars, suggesting that hot
Jupiters are not preferentially formed in these systems.
Then, if most hot Jupiters are not formed by the KL
mechanism in stellar binaries within $\sim100$ AU, 
we expect  that only a small fraction of the warm Jupiters 
can be explained by this channel because our numerical 
experiments indicate that the rate of formation of hot Jupiters is
higher that of warm Jupiters
(a factor of $\gtrsim4$ for binaries, see
below), while the observations indicate the opposite---there
are twice as many warm Jupiters than hot Jupiters.
On the other hand, \citet{wang15} recently found evidence that the
stellar multiplicity fraction of companions at
$\sim20-200$ AU is a factor of $\sim2$ higher for stars hosting a 
gas giant planet candidate from the Kepler sample, 
compared to a control sample with no planet detections.
\end{itemize}

We have carried out a population synthesis study
similar to that in Figure \ref{fig:pop}, but changing
the semimajor axis and mass ranges of the perturber from 
$a\out=5$--$6$ AU and $m\out=1-5M_J$ to
  $a\out=35$--$50$ AU and  $m\out=1M_\odot$, so the
the amplitude of the quadrupole potential $\phi_0$
in Equation (\ref{eq:phi_0}) lies roughly in the same
range. 
We observe that the ratio of hot
Jupiters to warm Jupiters formed in these simulations is 
$\simeq4$, compared to
$\simeq2$ in the planetary case. 
We also repeated the experiment with a broader eccentricity
distribution---uniform in $e\out^2$ over $[0,1]$ compared to $[0,0.3]$, which is 
more appropriate  for stellar 
binaries\footnote{We discard the systems that do not satisfy the 
stability criterion of \citet{MA01}.}.
As expected from our previous discussion, 
we find that the ratio of hot Jupiters to warm Jupiters
increases, from $\simeq4$ to $\simeq 5.2$.
Finally, we checked that the eccentricity distributions 
of the warm Jupiters from these experiments are
consistent with those from
our simple models in \S\ref{sec:ss_e}.

In conclusion, the eccentricity distribution of the warm
Jupiters predicted by our model is approximately flat regardless
of the whether the outer perturber is a planet or a star.
However, the ratio of warm Jupiters to hot Jupiter
is lower (by a factor $\sim 2$) in the binary-companion 
case compared to the planetary-companion case, making
the latter channel a more likely explanation of
the warm Jupiters.

\subsection{The ergodic approximation}

In \S\ref{sec:e_ergodic} we have explored 
a new analytical model to approximate
the time-averaged eccentricity distribution
of the inner binary in a hierarchical triple system.
This model is based on the ergodic hypothesis,
in which we assume that the planetary orbits randomly 
populate all the available phase space allowed by 
conservation of energy.  

We have shown that this model can reproduce 
the eccentricity and momentum coordinate 
$\tilde H\in=(1-e\in^2)^{1/2}\cos i\in$
distributions obtained from numerical three-body 
integrations when the inner orbit populates a significant
fraction of its available phase-space 
(see Figure \ref{fig:compa55}). 
This last condition is satisfied when there is 
a strong octupole-level gravitational 
perturbation from the outer orbit. 
For simplicity our discussion assumed that the inner body
is a test particle, but the model can be extended to
massive inner bodies.

The ergodic hypothesis is motivated by the observation that a
distribution function that is uniform on the energy surface in a
canonical phase space is always a solution of the collisionless
Boltzmann equation if the motion is governed by a Hamiltonian
\citep[e.g.,][]{bt08}. Thus we expect the distribution of systems over
the energy surface to be uniform if either the initial conditions
sample the phase space uniformly or the motion is sufficiently irregular
that the trajectory samples most of the phase space.

A more speculative application of the ergodic hypothesis to planetary
systems is described by \citet{T15}.

\section{Conclusions}

We have studied the steady-state orbital distribution of 
giant planets migrating through the combination of
secular gravitational perturbations due to a 
planetary or stellar companion and 
friction due to tides raised on the planet by the 
host star (``high-eccentricity migration'').

We have shown both analytically and numerically that the eccentricity
distribution of warm Jupiters arising from this migration mechanism is
approximately flat.  This distribution is inconsistent with the
observed eccentricity distribution of all of the warm Jupiters, which
decays approximately linearly from $e=0$ to $e=1$ (Fig.\
\ref{fig:hist}), but roughly consistent with the eccentricity distribution of
warm Jupiters with detected outer planetary companions, such as would
be required for high-eccentricity migration to occur. 

Both the observed eccentricity distribution and the observed ratio of
hot Jupiters to warm Jupiters are consistent with a model in which 
$\sim 20\%$ of warm Jupiters and 
most of the warm Jupiters with eccentricity $\gtrsim 0.4$ are
produced by high-eccentricity migration.

We also find that high-eccentricity migration
induced by a distant planetary 
companion can account for
the semimajor axes, the stellar obliquities, and occurrence rates of
most of the hot Jupiters.

Thus we are led to a model in which (i) high-eccentricity migration
produces most of the hot Jupiters; (ii) high-eccentricity migration
produces the $\sim20\%$ of warm Jupiters with $e\gtrsim 0.4$; (iii)
most of the remaining population of low-eccentricity warm Jupiters
must be accounted for by a different mechanism.

We also provide predictions for the mutual inclinations, spin-orbit
angles, and other properties of the hot and warm Jupiters produced by
high-eccentricity migration that can be used to test this model. 

\acknowledgements{
We are grateful to Bekki Dawson, Chelsea Huang,
Daniel Tamayo, Diego Mu\~noz, Marta Bryan,
Roman Rafikov, and Yanqin Wu
for enlightening discussions, and
the anonymous reviewer for valuable comments and
suggestions that improved the quality of the paper.
C.P. acknowledges 
support from the CONICYT Bicentennial Becas Chile fellowship, 
the Gruber Foundation Fellowship and
the Centre for Planetary Sciences at the 
University of Toronto.}


\appendix
\section{Time-averaged eccentricity distribution 
in a quadrupole potential}
\label{sec:appendix}

We calculate the time-averaged eccentricity 
distribution for the orbit of an inner test particle 
that evolves due to the quadrupole potential from
an external body.
We work in a coordinate system centered on the host star, of
mass $m_s$, in which the equator coincides with the orbit of the outer
body. We use Delaunay elements $L\in=(Gm_s a\in)^{1/2}$,
$G\in=L\in(1-{e\in}^2)^{1/2}$,  $H\in=G\in\cos i\in$; their conjugate angles
$\ell\in$, $\omega\in$, $\Omega\in$ are respectively the mean anomaly,
argument of pericenter, and longitude of the node of the test
particle. 

The doubly time-averaged Hamiltonian
that represents the gravitational potential of the outer body
up to quadrupole order is obtained by converting equation
(\ref{eq:phi_oct}) to orbital elements and dropping terms of order
$\tilde\epsilon_{\rm oct}$:
\ba
\mathcal{H}_{\rm q}=-\frac{\phi_0}{6(1-e\out^2)^{3/2}} 
\left[2+3e\in^2-(3-3e\in^2+15e\in^2\sin^2
  \omega\in ) \sin^2i\in \right].
\label{eq:ham}
\ea
where $\phi_0$ is given by Equation (\ref{eq:phi_0}).  Because the 
Hamiltonian is independent of $\ell\in$ and $\Omega\in$ the conjugate
momenta $L\in$ and $H\in$ are conserved (physically, this is because of
the secular approximation and because the quadrupole potential is
axisymmetric). Since the Hamiltonian is time-independent, the energy is also
conserved. Thus the test particle has only one degree of freedom. 

Let us fix the initial conditions by setting $e\in=e_0,$ $i\in=i_0,$ 
and $ \omega\in=\omega_0$. Then the component of
angular momentum normal to the outer orbit and the energy are fixed at 
\begin{equation}
H_0= L\in(1-e_0^2)^{1/2}\cos i_0\qquad \mathcal{H}_{{\rm q},0}=
-\frac{\phi_0}{6(1-e\out^2)^{3/2}} 
\left(2+3\theta_0\right), \mbox{ where }
\theta_0\equiv e_{0}^2-(1-e_{0}^2+5e_{0}^2\sin^2 \omega_{0} ) \sin^2i_{0}.
\label{eq:conserved}
\end{equation}

Because the motion has only one degree of freedom we can write the
steady-state phase-space distribution function as 
\ba
f(\omega\in,G\in)&\propto&
\delta\left(\mathcal{H}_{\rm q}-\mathcal{H}_{\rm q,0}\right) \nonumber\\
&\propto&\delta\left[ e\in^2-(1-e\in^2+
5e\in^2\sin^2 \omega\in ) \big(1-H_0^2/G\in^2\big) -\theta_0\right].
\ea
Then we can express the time-averaged eccentricity distribution as
\ba
n_e(e\in|\theta_0,H_0)&=&
\int\!\!\int
d\omega\in dG\in \,
f(\omega\in,G\in)\delta\Big( e\in-\sqrt{1-G\in^2/L\in^2}\,\Big),\\
&\propto&\int\!\!\int d\omega\in dG\in \,
\delta\left[ e\in^2 - (1-e\in^2+5e\in^2\sin^2 \omega\in ) 
\big(1-H_0^2/G\in^2\big)  -\theta_0\right]
\delta\Big( e\in-\sqrt{1-G\in^2/L\in^2}\Big),
\ea
and integrating over $G\in$ we get
\ba
n_e(e\in|\theta_0,H_0)&\propto&\frac{e\in}{(1-e\in^2)^{1/2}}\int d\omega\in\,
\delta\left[ e\in^2 - (1-e\in^2+5e\in^2\sin^2 \omega\in ) 
\left(1-\frac{H_0^2/L\in^2}{1-e\in^2}\right)  -\theta_0\right]
\nonumber\\
&\propto&
\frac{ e\in}{
\big\{\big[2e\in^2+H_0^2/L\in^2-\theta_0-1 \big]
\big[\left(1+4e\in^2\right)\left(1-e\in^2-H_0^2 /L\in^2\right)
-\left(e\in^2-\theta_0\right)\left(1-e\in^2\right)
\big]\big\}^{1/2} 
  }, \nonumber \\
  \label{eq:n_e}
\ea
or zero if the argument of the square root is not
positive. The normalization is chosen so 
$\int_{0}^{1} de\in n_e(e\in|\theta_0,H_0)=1$.

\subsection{Migrating planet}

A migrating planet must have $e\in\simeq 1$ at some point on its
trajectory. If we define the initial conditions at this point then 
we can express the initial energy parameter as
\ba
\theta_0=1-5\sin^2\omega_0\sin^2i_0=1-5\cos^2\psi_0.
\label{eq:theta_psi}
\ea
where $\psi_0$ is the polar angle of the eccentricity vector at the
initial time, determined by $\cos \psi\equiv {\bf \hat{e}}\in\cdot
{\bf \hat{h}}\out$ (see footnote \ref{footpsi}). Moreover if
$e\in=1$ at any point on the trajectory then the conserved
quantity $H\in=0$. Then the time-averaged eccentricity distribution in  Equation
(\ref{eq:n_e}) becomes
\ba
n_e(e\in|\psi)&\propto& \frac{e\in}{
\left(2e\in^2-2+5\cos^2\psi_0 \right)^{1/2}(1-e\in^2)^{1/2}
\left(2+3e\in^2-5\cos^2\psi_0\right)^{1/2}   }.
  \label{eq:n_e_psi}
\ea


\end{document}